# Interaction dust – plasma in Titan's ionosphere: feedbacks on the gas phase composition

**Audrey Chatain** [a,b,*,1], **Nathalie Carrasco** [a], **Ludovic Vettier** [a], **and Olivier Guaitella** [b]

[a]LATMOS, Université Paris-Saclay, UVSQ, Sorbonne Université, CNRS, Guyancourt, France

[b]LPP, Ecole polytechnique, Sorbonne Université, Institut Polytechnique de Paris, CNRS, Palaiseau, France

* Corresponding author: audrey.chatain@latmos.ipsl.fr

[1]Present affiliation: Departamento de Física Aplicada, Escuela de Ingeniería de Bilbao, Universidad del País Vasco/Euskal Herriko Unibertsitatea (UPV/EHU), Bilbao, Spain.

## Highlights

• Titan's organic aerosols interact with the ionospheric $N_2$-$CH_4$-$H_2$ plasma species

• $N_2$-$H_2$ plasma species are especially efficient to erode organic surfaces

• Analogues of Titan's aerosols are exposed to a $N_2$-$H_2$ plasma

• Newly formed HCN, R-CN and carbon-bearing ions are observed by mass spectrometry

• Suggested processes: surface chemistry by radicals and desorption by ion sputtering

## Abstract

Titan's organic aerosols are formed and spend some time in the ionosphere, an atmospheric layer ionized by solar VUV photons and energetic particles coming from the magnetosphere of Saturn, forming a natural $N_2$-$CH_4$-$H_2$ plasma. Previous works showed some chemical evolution processes: VUV photons slightly alter the aerosols nitrile bands, hydrogen atoms tend to hydrogenate their surface and carbon-containing species participate to the growth of the aerosols. This work investigates the effect of the other plasma species, neglected until now, namely the $N_2$-$H_2$ derived ions, radicals and electronic or vibrational excited states. Industrial plasmas often use $N_2$-$H_2$ discharges to form ammonia-based fertilizers, for metal nitriding, and to erode organic surfaces. Consequently, these are likely to affect Titan's organic aerosols.

We therefore developed the THETIS experiment (for Tholin Evolution in Titan's Ionosphere Simulation) to study the interactions between analogues of Titan's aerosols (tholins) and the erosive $N_2$-$H_2$ plasma species found in Titan's ionosphere. Following a first paper on the investigation of the evolution of the solid phase by Scanning Electron Microscopy and IR transmission spectroscopy (Chatain et al., 2020a), this second paper focuses on evolution of the gas phase composition, studied by neutral and ion mass spectrometry.

In the presented experiments, newly formed HCN, $NH_3$-CN and $C_2N_2$ are extracted in quantity from the tholins as well as some other carbon-containing species and their derived ions. On the other hand, the production of ammonia strongly decreases, probably because the H, NH and N radicals are rather used for the production of HCN at the surface of tholins. Heterogeneous processes are suggested based on the experimental observations: chemical processes induced by radicals at the surface would modify and weaken the tholin structure, while ion sputtering would desorb small molecules and highly unsaturated ions. The effect of plasma erosion on aerosols in Titan's ionosphere could therefore lead to the formation of C≡N bonds in the aerosol structure, as well as the production of HCN or R-CN species in the gas phase.

## Graphical abstract

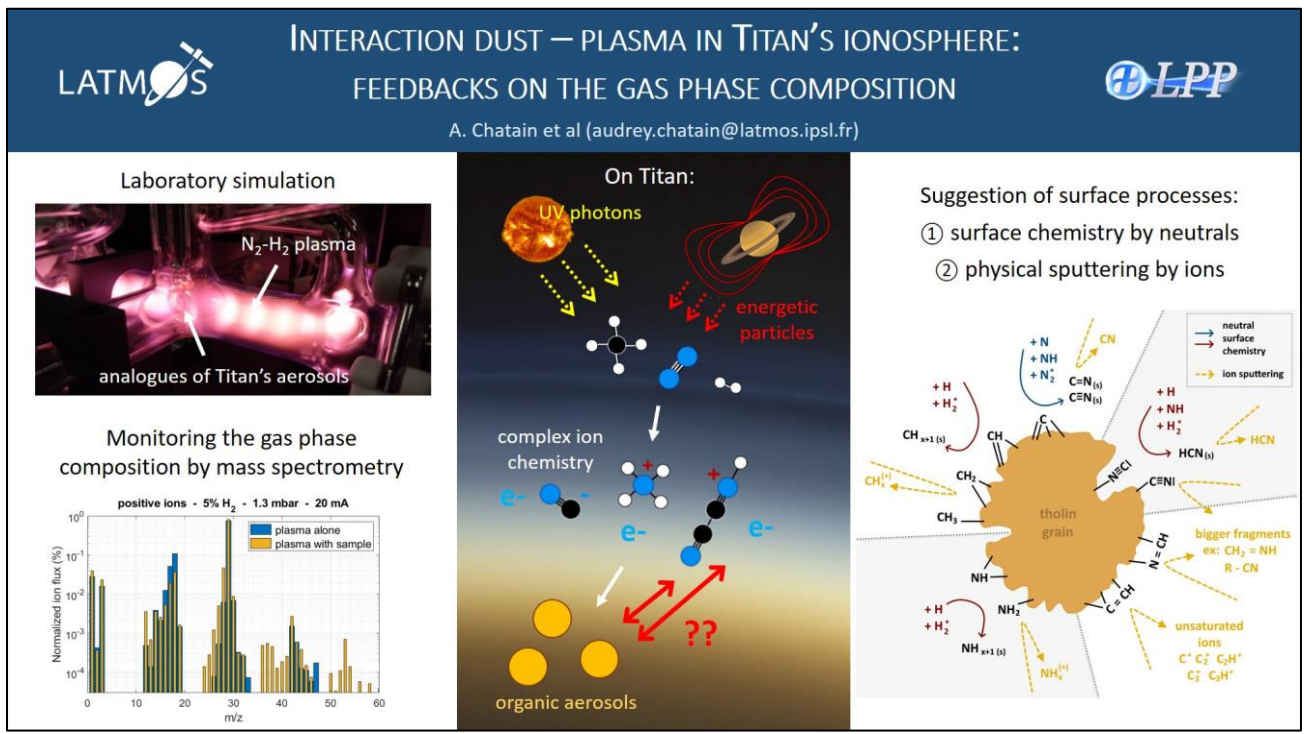

## 1- Introduction

Titan's organic aerosols reside several days in the ionospheric layer ionized by VUV radiations and magnetospheric electrons from 1200 to 900 km (Lavvas et al., 2013), and are exposed to a $N_2$-$CH_4$-$H_2$ natural plasma during this stay. The effect of such an exposure is still unknown, but strong heterogeneous interactions are expected and some have been reproduced and observed in the laboratory (Chatain et al., 2021).

There are several examples of organic matter in astrophysics being transformed by exposure to various irradiations. Organic ices present in protoplanetary discs are exposed to extreme conditions (radiations, temperature, energetic particles…) and evolve into complex organic matter (Fray et al., 2016; Herd et al., 2011). Laboratory studies have already been performed to mimic a possible evolution of organic matter in protoplanetary discs, in the interstellar medium and on comets. Depending on the energy deposited, the exposure of an organic material to ion, electron and photon (X rays and UV) irradiation lead to changes in morphology and/or in chemical composition (Öberg, 2016).

The combined effects of ion, electron and photon irradiations are processes similar to what would occur in a plasma and in particular in Titan's ionosphere. The evolution of laboratory analogues of Titan's aerosols (named 'tholins') exposed to VUV radiations has been previously investigated by Carrasco et al. (2018). They observed modifications in the nitrile infrared bands of tholins. The interaction of atomic hydrogen with tholins has been studied in Sekine et al. (2008). They observed some hydrogenation of the tholins and recombination of $H_2$. The next logical step is then to investigate the complementary effect of electrons, ions, radicals and electronic or vibrational excited states on Titan's organic aerosols.

Organic matter is highly sensitive to plasma species. There are a lot of examples in the industry where organics are eroded and/or transformed by plasma exposure (Bogaerts et al., 2002). Cold plasma discharges are used to change the properties of surfaces: coating for protection or strength, surface sterilization or modification of properties such as the wettability, printability, bio-compatibility, the thermal and electrical conductivities… They also induce etching on specific surfaces, especially used in the microelectronics and semiconductor fields. These properties of the plasma to induce surface processing make them essential in many sectors: biomedical, energy research, environmental… Plasmas containing carbon (e.g. $CH_4$) generally lead to carbon chain growth, forming films and/or particles. On the opposite, $N_2$-$H_2$ plasmas are often used to remove organic matter. On Titan, we expect both the carbon-growth (led by the presence of methane) and the $N_2$-$H_2$ erosion processes to be in competition. The complete system is extremely complex. In the objective to identify individual mechanisms at work, our strategy is to go step by step. In this work we first focus on the erosion processes induced by $N_2$-$H_2$ plasma species. The carbon-growth processes and their competition with $N_2$-$H_2$ erosion processes are planned for future studies.

A first example of the erosion induced by $N_2$-$H_2$ plasmas is the cleaning of organic films deposited on walls of future nuclear fusion devices (Voitsenya et al., 2006). They observe that a higher etching rate is obtained if a heavy element is added to $H_2$, like $N_2$. Plasmas in $N_2$ and/or $H_2$ and/or Ar are also widely used for organic film etching. Studies are especially focused on diverse polymers, and on low-dielectric constant films used in integrated circuits. Both physical and chemical modifications are observed (Kurihara et al., 2005). The etching of the organic films also leads to the formation of new gaseous species. Hong et al. (2002), Hong and Turban (1999), Ishikawa et al. (2006) and Kurihara et al. (2006) observed the decomposition of the films exposed to a $N_2$ or $N_2$-$H_2$ plasma into HCN, CN and $C_2N_2$. Van Laer et al. (2013) modelled the processes at work in the erosion of a (C:H) organic film by a $N_2$-$H_2$ plasma. They predict the formation of many neutral species (from highest to lowest density: N, H, N*, NH, $NH_3$, HCN, $CH_4$, $NH_2$, $CH_3$, CN, CH, C, $CH_2$) and ions ($N_2H^+$, $N_2^+$, $H_2^+$, $NH^+$, $N^+$, $H^+$, $NH_2^+$, $NH_3^+$, $H_3^+$, $NH_4^+$, $CH_3^+$, $CH_4^+$, $HCN^+$, $CH_2^+$, $CH^+$, $C^+$).

Heterogeneous processes have been investigated in detail in these plasma etching experiments. Depending on the plasma conditions, three main processes are expected: the sputtering of energetic ions, the reaction with short life excited species of high chemical reactivity (radicals, ions, vibrationally and electronically excited molecules) and the interaction with VUV photons emitted by the hydrogen plasma (Moon et al., 2010; Uchida et al., 2008). Nagai et al. (2002) studied the importance of H radicals for the etching. They observed that on the opposite, N radicals tend to form a protection layer against the etching. N radicals and $N_2$ interact with the organic film to form CN. The addition of $N_2$ in $H_2$ plasmas forms erosive $N_2H^+$ ions, but also protective N radicals that could passivate the surface. They found a maximum etching rate for 30% $N_2$ in $H_2$. Jacob et al. (2006) suggest that atomic hydrogen and ions could have a combined effect. Moon et al. (2010) also include the ammonium ions $NH_3^+$ and $NH_4^+$ in the etching process. Wertheimer et al. (1999) performed different experiments to irradiate various polymers to VUV photons produced by plasma. They observed some evolutions, but far less efficient than the direct effect of ions and radicals in a plasma.

In conclusion, $N_2$ and/or $H_2$ plasma species strongly interact with organic films and polymers, both physically and chemically. Studies have been performed in different gas mixtures, pressure and temperature conditions, and on various organic substrates. In this work we investigate what can be expected for the organic aerosols in Titan's ionospheric conditions.

Our objective is to analyze the interaction between the aerosols and the non-carboneous plasma species in the ionosphere of Titan. For this purpose, we expose analogues of Titan's aerosols ('tholins') to a plasma representative of the ionosphere of Titan in the THETIS experiment (for Tholin Evolution in Titan's Ionosphere Simulation). The setup is dedicated to the study of the exposure to a $N_2$-$H_2$ plasma, the erosive part of Titan's ionospheric plasma. A first paper (Chatain et al., 2020a) has already investigated the evolution of the solid phase by scanning electron microscopy and infrared transmission spectrometry through the tholin sample. A strong erosion of the grains was observed, with the appearance of nanostructures in the previously round and smooth grains of 200-300 nm, and the modification of chemical functions (especially nitriles). The strong morphological and chemical evolution of the solid phase suggests a possible release of molecules and ions into the gas phase. This feedback of the erosion process on the plasma chemical composition is the aim of the present paper. The neutral molecules and positive ions are monitored during the experiment by mass spectrometry. We observe the formation of new species and the disappearance of others. The observations are interpreted in the light of the results on the solid phase (Chatain et al., 2020a) and previous works on plasmas etching experiments. This work is a joint study with Chatain et al. (n.d.), in which the $N_2$-$H_2$ plasma discharge (without tholins) is thoroughly studied.

## 2- Experimental setup

### 2.1- The THETIS experiment

The experimental basis is the same as described in Chatain et al. (2020a), which should be referred to for more experimental details. First, analogues of Titan's aerosols ('tholins') are produced using the PAMPRE experiment (Szopa et al., 2006), in its standard working conditions (Sciamma-O'Brien et al., 2010). Round grains of 200-300 nm are formed. After production, tholins are temporarily exposed to ambient air during their collection and the making of pellets. To make a pellet, 1 mg of tholins is homogeneously spread using a spatula and an absorbent napkin on a stainless steel grid of 1.3 cm in diameter, with 25 µm-large threads and 38 µm-wide meshes. When exposed to air, water adsorption and oxidation appear on the first nanometers of the grains, as measured in Carrasco et al. (2016). To prevent the further evolution of the samples with air (Maillard et al., n.d.), those are stored under vacuum and heated at 80-100°C to desorb water.

The geometry of the THETIS reactor is not conventional and has been specifically designed for the experiment. The reactor is formed with two perpendicular pyrex tubes, with inner diameters of 2 cm and openings at the end of each tube (see Figure 1b). One tube has hollow cylindrical iron electrodes coated with nickel with a small quantity of barium fluoride ($BaF_2$) positioned perpendicularly at each extremity and is dedicated to the plasma discharge. The second axis is used to couple the instruments. A great care is also taken in cleaning the inside of the reactor before each experiment. If previous experiments with the reactor involved organic compounds, the inside of the glass tube is first manually cleaned with ethanol – and sometimes diluted acetic acid. Then the tube is closed and pumped to high vacuum (~$10^{-6}$ mbar) with a turbo-molecular pump. To remove compounds – especially water – adsorbed on the walls, the tube is heated to ~80°C with a heating cord and aluminum foil for 50 min (though we suspect that some water still remains on the metallic electrodes). Finally, the tube is let under vacuum during > 30 min, until the mass spectrometer detects very few impurities.

During the experiment, we expose the samples in the positive column of a DC glow discharge, also used in Chatain et al. (2020a) and Chatain et al. (n.d.). The plasma is ignited in one of the pyrex tubes long of 23 cm. A gas mixture of $N_2$-$H_2$ at low density (0.5 to 5 mbar) is continuously injected in the tube. The gas circulation in the reactor is provided by four long pyrex tubes of 10 mm in internal diameter connected to the four extremities of the reactor, perpendicular to the main tubes. These long tubes are coiled to gain space. Their length is necessary to avoid the discharge to start on any grounded metallic piece in the gas circuit outside the reactor. Two opposite ones serve for gas input and the two others for gas output. We use high purity gases (> 99.999%). The gas from a 95% $N_2$ – 5% $H_2$ gas bottle (Air Liquide – CRYSTAL mixture) is diluted in pure $N_2$ (Air Liquide – alphagaz 2). The mixing is obtained by two mass flow controllers (Brooks 5 standard cubic centimeter per minute (sccm), full scale accuracy 1%). The total flow is fixed at 5 sccm. The ignition of the discharge inside the tube is triggered by the application of a 1-3 kV voltage between the electrodes. Currents studied are 10-40 mA. The reference study uses a current of 20 mA at 1 mbar, which gives an electron density in the tube of about $3.10^9$ cm$^{-3}$ and a gas temperature of ~360 K (Brovikova and Galiaskarov, 2001; Chatain et al., n.d.). We note that both for the highest current case (40 mA, 1 mbar), and for the highest pressure case (20 mA, 4 mbar), the gas temperature can reach 400 K. Depending on the experimental conditions, the reduced electric field $E/N_g$ varies between $3\ 10^{-16}$ and $10\ 10^{-16}$ V.cm² (Chatain et al., n.d.). The corresponding electron energy distribution function (EEDF) can be estimated from the figure 1 of Guerra et al. (2004). Chatain et al. (2020a) has investigated the validity of the simulation of ionospheric conditions in the laboratory and concluded that erosion processes are faster in our experimental simulation than on Titan. In terms of surface erosion, a week on Titan would correspond to a few seconds in the laboratory. However, laboratory grains are more than ten times larger than the aerosols in Titan's atmosphere. As a result, the erosion of one grain takes more time. For this reason, we investigated the processes happening over longer periods of time.

The sample is deposited on a movable glass stick that has a circular shape at its extremity perfectly adjusted to the pellet size and that can be moved under vacuum up to 30 cm away from the plasma glow, in a perpendicular vertical axis (see Figure 1a). The position far from the plasma is used to ignite, stabilize and analyze the plasma discharge before the insertion of the sample (Chatain et al., n.d.).

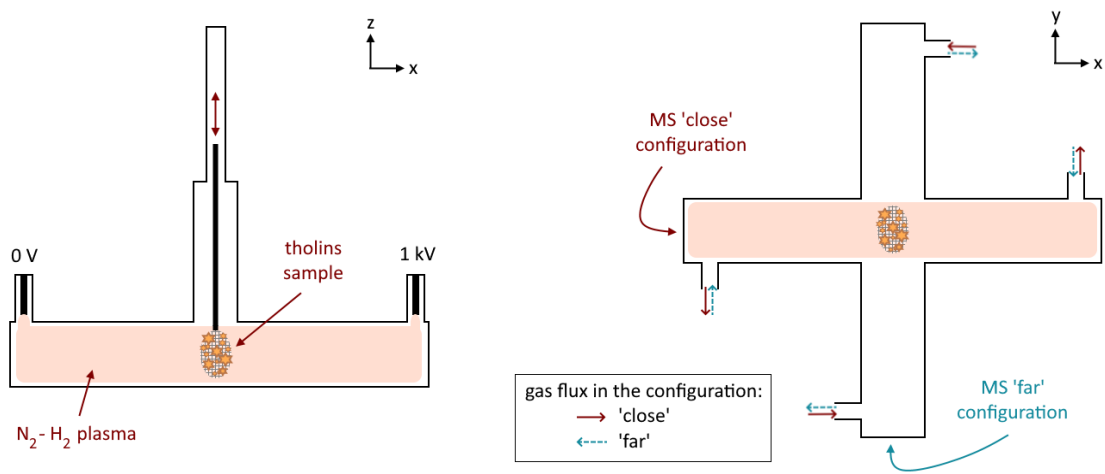

**Figure 1: Scheme of the experimental setup THETIS for the analysis of the gas phase.**

## 2.2- Analysis of neutral and ionized gas species by mass spectrometry

Neutral molecules and positive ions are investigated thanks to a quadrupole mass spectrometer (a Hiden analytical, electrostatic quadrupole plasma (EQP) series, further abbreviated 'MS', and previously used in Chatain et al., 2020b). For the ion measurements, the collecting head of the mass spectrometer required to be only a few centimeters from the plasma, close to the grounded electrode (see the 'close' configuration in Figure 1b). However, to measure the perturbations that the collecting head could induce on the plasma by being close to it, neutral measurements have been performed in two configurations: close to the plasma (at a few centimeters like for the ions) and far from the plasma (at ~30 cm, see the 'far' configuration in Figure 1b). To maximize the collection of newly formed species, the pumping system is always positioned in the same branch as the MS.

## 3- Disappearance and formation of neutral species

### 3.1- Acquisition method and general observations

*3.1.1- An acquisition in continuous mode*

The mass spectrometer continuously takes spectra during the experiment. Each spectrum is from 1 to 100 u with a resolution of 1 u, and the acquisition takes ~30 s. Figure 2 shows the evolution of a few different m/z in the case of an experiment in which tholins are exposed to a plasma in $N_2$-$H_2$ with 5% $H_2$, at 1.3 mbar and with a current of 20 mA. These conditions will be referred to as the reference conditions in the following parts.

In this case, the gases $N_2$ (at 28 u) and $H_2$ (at 2 u) are injected in the experiment at t = 6 min. Then, water that comes from impurities in the gas bottle and/or desorption from the reactor electrodes can be detected at 18 and 17 u (17 u being due to OH formed by fragmentation of $H_2O$ in the MS). Once the pressure is stabilized, the plasma is ignited in the reactor (at t = 13 min). As discussed in Chatain et al. (n.d.), plasma discharges in $N_2$-$H_2$ mixtures lead to the formation of ammonia (observed at 17 u). The higher relative increase of 17 u compared to 18 u proves that this strong increase is mostly due to $NH_3$, and not to the OH fragment of $H_2O$. The ignition of the plasma also helps to desorb water inside the reactor (seen at 18 u). During this phase, the tholin sample is positioned ~30 cm away from the plasma. Nevertheless, we suspect that some plasma species could interact with the tholins at this distance, participating in the formation of traces of HCN (27 u) and other carbon-containing compounds (like 41 u) detected since the ignition of the plasma. However, Chatain et al. (n.d.) shows that this contribution is minor.

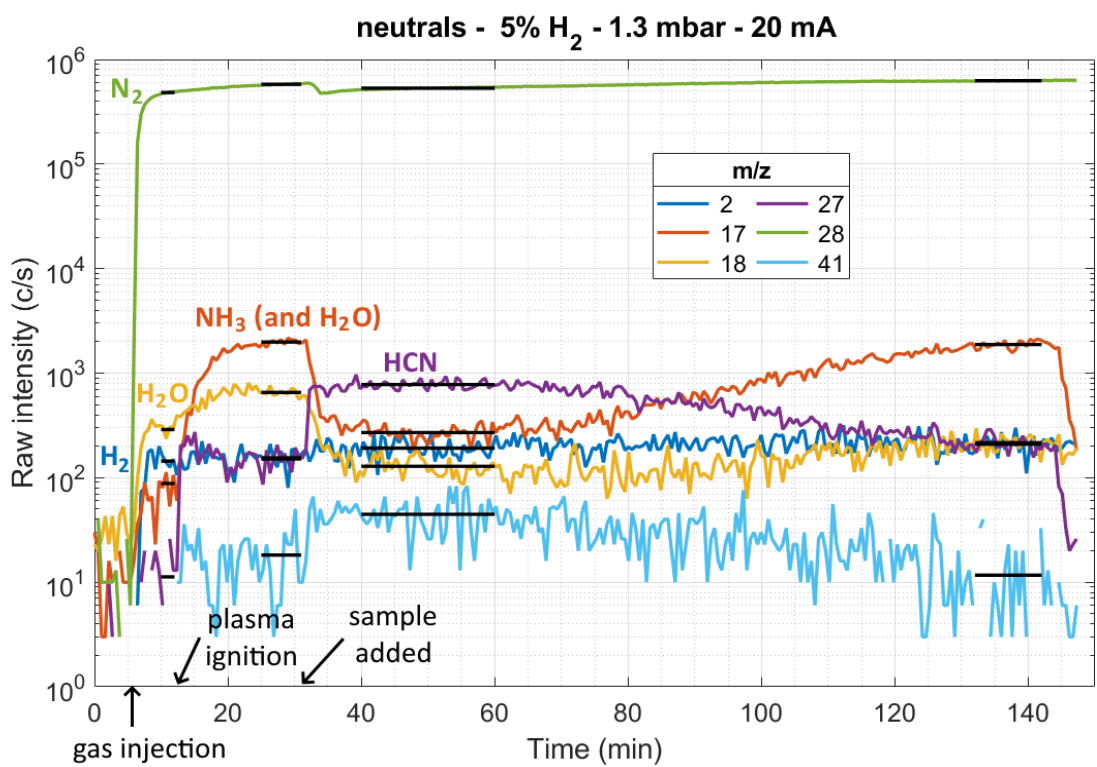

**Figure 2:** MS measurement of a few masses during a complete experiment of tholin exposure to a $N_2$-$H_2$ plasma (5% $H_2$, 1.3 mbar and 20 mA). The gas mixture is injected at t = 6 min, the plasma is ignited at t = 13 min, the sample is added at t = 32 min and the plasma is stopped at t = 145 min. The black lines represent the stabilized values for each mass on the different sections.

The formation of ammonia takes a few minutes to stabilize, then the sample is introduced in the plasma (at t = 32 min). At this point, an important modification of the gas phase composition is observed. The ammonia formed in the $N_2$-$H_2$ plasma alone suddenly drops (decreased by 85% in the conditions of Figure 2). On the opposite, new carbon-containing species are formed (like 41 u). The main neutral species, aside from $N_2$ and $H_2$, becomes HCN. After 130 min, the neutral gas composition is identical to the $N_2$-$H_2$ plasma conditions without tholins. Indeed, the tholins on the grid have all been eroded, possibly reduced to small fragments, molecules and ions dragged by the low gas flow, and nothing remains on the metallic grid. The exposure time needed to 'clean' the grid from tholins depends on the plasma parameters chosen and on the exact spreading of tholins on the pellet. In the conditions tested, it varied from 40 min (at 40 mA) up to more than 3h (time limit after which the experiment is stopped). This also shows that the presence of the sole metallic grid in the plasma does not affect the neutral species.

*3.1.2- Average and calibration of the signal*

Figure 2 directly shows the intensity measured by the MS. Noisy variations are observed, especially for species at low intensity. In the following we therefore consider average values taken on 4 specific zones: (1) the gas mixture before plasma ignition, (2) the plasma without sample and (3)-(4) the plasma with sample at the beginning and at the end of the experiment. The average values chosen in the case of Figure 2 are plotted in black lines. Except for (4), these zones are approximately at the same time periods in the different experiments performed.

As discussed in the Appendix D of Chatain et al (2020b) that used the same MS and to refer to for more details, the raw intensity measured by the MS has to be corrected by a mass-dependent transmission function. Then, the deduced intensity is in Pa.Å². To obtain the partial pressure (or the number density) of the species in the gas phase, the fragmentation pattern, the isotopes and the ionization cross section (Å²) of the different species have also been taken into account. The MS transmission of low masses (m/z ≤ 4) is very different from the transmission of the other masses. To maximize the signal, low masses have usually been acquired with MS parameters tuned on m/z 2, while the other masses have been measured with MS parameters tuned on m/z 28. The transmission curves used to correct both mass regions are therefore different (see Figure D2 in Chatain et al., 2020b). In both cases, the transmission curve is a log-normal law (see Equation [1]). For the tuning focused on m/z 2, the coefficients are: $G = 37\,100\,;\mu = 1\,;s = 1.57$. And for the tuning focused on m/z 28, we used: $G = 290\,000\,;\mu = 3.42\,;s = 1$.

$$f(x; G, \mu, s) = \frac{G}{x.\,s.\,\sqrt{2\pi}} \cdot \exp\left(-\frac{(\ln(x) - \mu)^2}{2s^2}\right) \qquad [1]$$

### 3.1.3- Evolution of the spectrum in reference conditions

Figure 3 shows the spectra obtained in the reference conditions for the 3 first zones defined above: (1) gas mixture before the plasma ignition, (2) plasma without sample and (3) plasma with sample. The evolution of the main gas species is the same as observed on Figure 2. In addition, the complete spectra presented in Figure 3 show all the new organic species formed by exposure of the tholins to the $N_2$-$H_2$ plasma. A summary of the observations is given in Table 1.

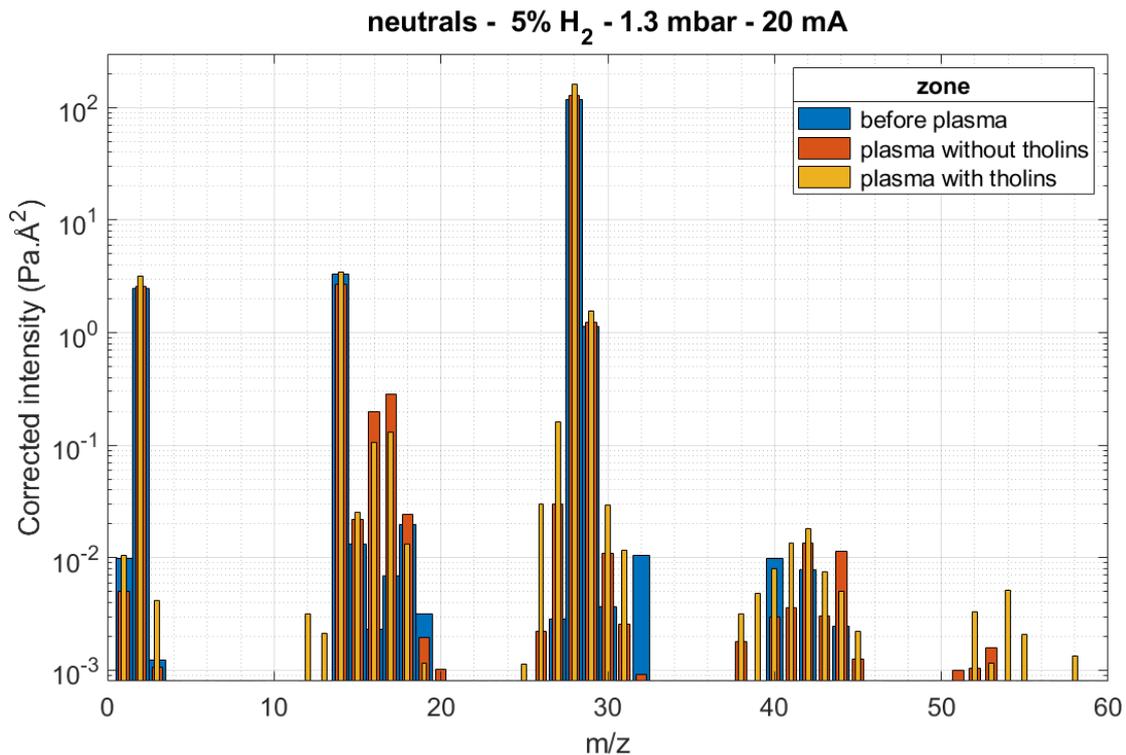

**Figure 3: Spectra obtained for the reference experiment (5% $H_2$, 1.3 mbar, 20 mA): before plasma (large blue bars), after ignition and stabilization of the plasma (medium orange bars), and after the addition of the tholins sample (thin yellow bars).**

|  | gas | plasma | sample |
|---|---|---|---|
| formation | injected:<br>**N$_2$**: m/z 14, 15, **28**, 29, 42<br>**H$_2$**: m/z 1, **2**<br><br>*traces:*<br>*H$_2$O: m/z 16, 17, **18***<br>*O$_2$: m/z 32*<br>*Ar: m/z 40*<br>*CO$_2$: m/z 44* | **NH$_3$**: m/z 16, **17**<br><br>*traces:*<br>*organics at m/z 26, 27, 30, 31, 38, 41-43*<br>*CO$_2$: m/z 44 (at the beginning, from O$_2$)* | **HCN**: m/z 26, **27**<br><br>**other carbon-bearing molecules**, with peaks at m/z 26-31, 38-45, 52, 54 |
| loss |  | *traces:*<br>*O$_2$, Ar, H$_2$O*<br>*(decrease with time)* | **NH$_3$** |

**Table 1: Summary of the modifications of the neutral gas phase species during the experiment in reference conditions. The threshold for the 'traces' in the gas and plasma cases is set to 3x10-2 Pa.Å² on Figure 3.**

We observe a non-common contribution at m/z 42, even present in the absence of plasma discharge. It is systematically detected when nitrogen is present in the gas phase: with only gases and with plasma, with the pure N$_2$ bottle or the N$_2$-H$_2$ bottle, in different reactors. Besides, its intensity is very reproducible. Junk and Svec (1958) had a similar observation and investigated its origin. They found that N$_3^+$ was formed inside the MS, by low probability endothermic reactions. This point does not impact our results, except at m/z 42 that we will not discuss further.

The formation of HCN (and maybe CN) at 26 and 27 u is observed, as well as the formation of C$_2$N$_2$ at 52 u. These gas products have also been detected by Optical Emission Spectroscopy in Hong and Turban (1999) and by mass spectrometry in Hong et al. (2002) and Kurihara et al. (2006) as they exposed various (C:H) polymers to a N$_2$ plasma. Ishikawa et al. (2006) explain that in a N$_2$-H$_2$ plasma, CN and NH bonds are formed at the surface of the organic film. In particular, CN bonds submitted to ion bombardment form CN, HCN and C$_2$N$_2$.

Many other gas species are observed in our experiment, in the case of tholins exposed to a N$_2$-H$_2$ plasma. They can be due to the presence of nitrogen and various chemical structures in the tholins, contrarily to the (C:H) polymer films. The main suspected molecule is acetonitrile CH$_3$-CN with fragments from 38 to 41 u**.** Other molecules are suggested in Table 2.

| m/z | suggestion of attribution | m/z | suggestion of attribution |
|---|---|---|---|
| 26 | fragment HCN<br>acetylene ($C_2H_2$)<br>radical CN<br>*fragment ethene*<br>*fragment ethane* | 40 | fragment acetonitrile |
| 27 | **hydrogen cyanide (HCN)**<br>*fragment ethene*<br>*fragment ethane* | 41 | **acetonitrile ($CH_3$-CN)**<br>*main fragment propene ($C_3H_6$)* |
| 28 | $N_2$<br>*fragment ethylamine*<br>*fragment dimethylamine*<br>*fragment propane nitrile*<br>*ethene ($C_2H_4$)*<br>*main fragment ethane ($C_2H_6$)* | 42 | MS artefact<br>*cyanamide ($NH_2$-CN)*<br>*main fragment ethylenimine*<br>*fragment propene* |
| 29 | $N_2$ isotope<br>*methanimine ($CH_2$=NH)*<br>*fragment ethane* | 43 | ethanimine ($CH_3$-CH=NH)*<br>ethenamine ($NH_2$-CH=$CH_2$)*<br>N-methylmethanimine ($CH_2$=N-$CH_3$)*<br>fragment ethylenimine (cyclic-$CH_2$(NH)$CH_2$) |
| 30 | $N_2$ isotope<br>main fragment methylamine ($CH_3$-$NH_2$)<br>*main fragment ethylamine ($CH_3$-$CH_2$-$NH_2$)*<br>*fragment ethane* | 44 | main fragment dimethylamine ($CH_3$-NH-$CH_3$)<br>*fragment ethylamine* |
| 31 | fragment **methylamine** | 45 | fragment dimethylamine<br>*fragment ethylamine* |
| 38 | fragment acetonitrile | 52 | **cyanogen (NC-CN)** |
| 39 | fragment acetonitrile<br>*fragment propene*<br>*main fragment butadiene ($C_4H_6$)* | 54 | propane nitrile ($C_2H_5$-CN)<br>*fragment butadiene* |

**Table 2: Suggestions of organic contributions of the MS peaks detected in Figure 3. The fragmentation patterns of the molecules come from the NIST database. The ones unknown are marked with a *.**

Further tests have been performed on this experiment to ensure the robustness of the results. We sum up our conclusions here and give more details in the appendix.

First, the experiment presented on Figure 3 has been repeated three times with exactly the same conditions and on different weeks. Results are highly repeatable (as shown on Figure A1-1 in Appendix A1.1).

Secondly, we investigated the effect of the MS collecting head position on the detected molecules. The results presented in Figure 3 were obtained with the MS collecting head aligned with the plasma axis, as shown in the 'close configuration' in Figure 1b. In this configuration, the plasma is only a few centimeters away from the MS. Then, the metallic collecting head is very close to the plasma and could induce some perturbations. To quantify this effect, two other experiments were conducted in the same reference conditions, but with the MS collecting head further from the plasma (at ~30 cm, 'far configuration' in Figure 1b). The results are globally identical. Some minor changes on the detection of ammonia, some amines and nitriles are discussed in Appendix A1.2. To validate the results presented further, experiments have been performed at least once in each position.

Third, to confirm the effect of tholins on the plasma, an experiment was done without tholins on the grid. In this case, no changes have been detected with the addition of the sample, except perhaps a very slight increase of ammonia (x2), as shown in Figure A1-4 in Appendix A1.3. The metallic grid certainly enhances the surface production of $NH_3$. In conclusion, all the effects observed previously are due to the presence of tholins.

## 3.2- Variations with the injected gas

In the previous sub-section, we observed that the exposure of tholins to a $N_2$-$H_2$ plasma leads to the formation of new carbon-containing species. As no carbon is injected in the gas phase, it necessarily comes from the tholins. To investigate the nature of the processes forming these new carbon-bearing molecules, we performed experiments with different gases.

First, to examine the effect of $H_2$ and $N_2$ separately, we exposed tholins to a pure $N_2$ plasma. The spectra obtained are given in Figure 4a. Nearly no modifications are observed in the spectrum with the ignition of the plasma and the addition of the sample. In particular, traces of $H_2O$ (18 u) are transformed into traces of $H_2$ (2 u) at the ignition of the plasma. In conclusion, the sole plasma species formed in a $N_2$ plasma do not affect tholins much compared to the $N_2$-$H_2$ plasma. The observations seen in the above section are thus not triggered only by plasma sputtering of pure $N_2$ plasma species. The presence of hydrogen in the plasma is necessary to erode the tholins, possibly through chemical reactions including H-bearing species.

In more details, Figure 4a shows a small increase at m/z 26, 27 and 52 attributed to HCN and $C_2N_2$. The formation of these species was also observed in Hong et al. (2002), Hong and Turban (1999) and Kurihara et al. (2006) with the exposure of (C:H) films to $N_2$ plasmas. Many works discuss the importance of the hydrogen in $N_2$ plasma discharge for an enhanced film erosion (Ishikawa et al., 2006; Morikawa et al., 2003; Nagai et al., 2003, 2002; Van Laer et al., 2013). They conclude on a synergistic effect induced by $N_2$ and $H_2$ plasma species. Nitrogen is added to the surface especially by the radicals N*, $NH_2$* and $(N_2)$*, then the ion bombardment leads to a release of H, forming the double and triple bonds C=C, C=N, C≡N, that finally lead to the stable products HCN and $C_2N_2$. In particular, HCN requires H* and *$NH_2$ to form from -C≡N. Consequently, its production is enhanced by the presence of hydrogen in the gas phase. Otherwise, the surface is passivated by an accumulation of -C≡N bonds, explaining the lower etching rates in pure $N_2$ plasma.

For comparison, an experiment was performed in a pure argon plasma. Except traces of $H_2$, no new molecules were formed at the ignition of the plasma and with the addition of the tholin sample (see Figure 4b). Consequently, the processes eroding tholins require the use of both hydrogen and nitrogen. Tholins cannot just be decomposed into smaller molecules by sole ion bombardment; a combined nitrogen and hydrogen chemistry is required. The experiment in argon also shows that the observed effects in a $N_2$-$H_2$ plasma are not just due to the increase in temperature of a few 10s of degrees induced by the plasma discharge. Indeed, this increase is nearly identical in all cases, but no modifications are observed in pure argon.

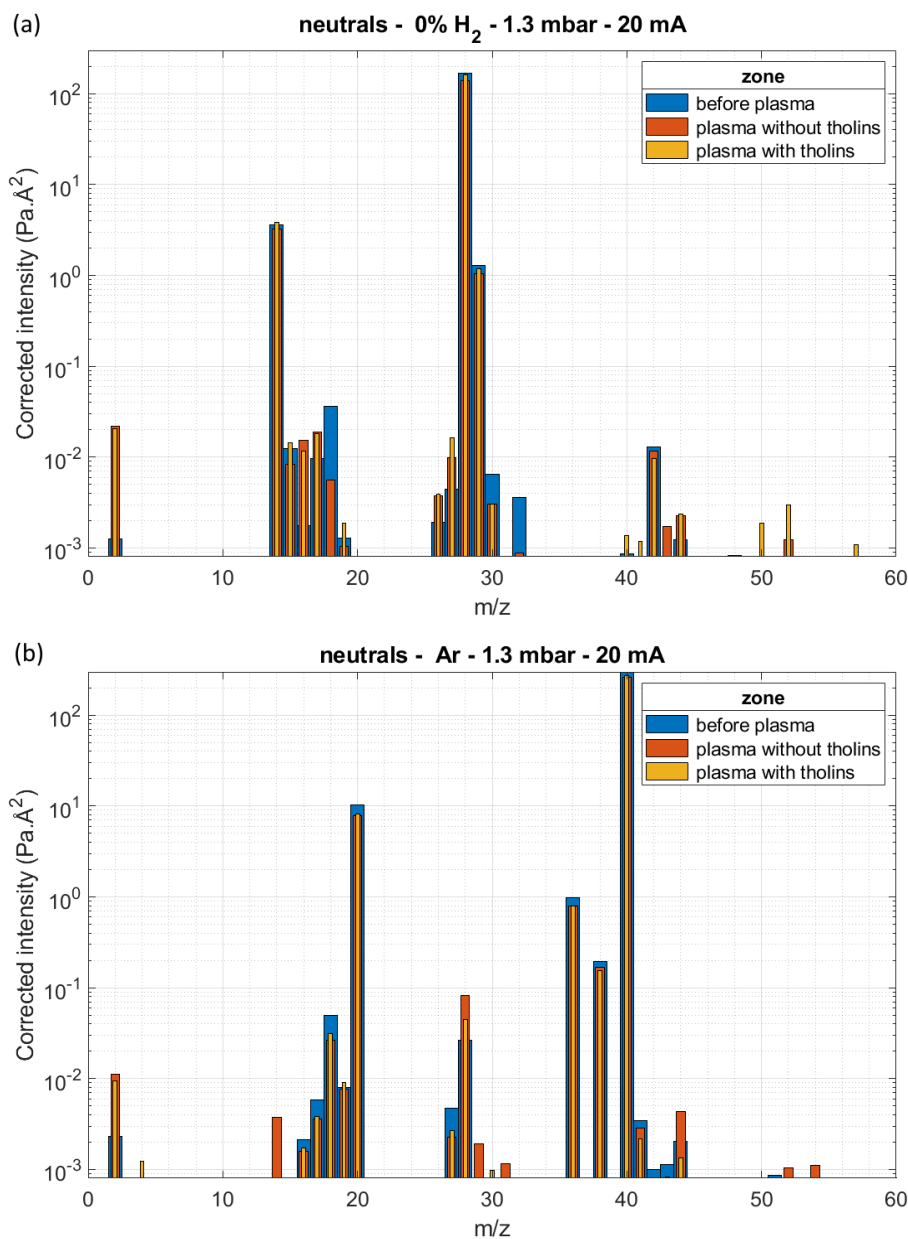

Figure 4: Spectra obtained before plasma (large blue bars), after ignition and stabilization of the plasma (medium orange bars), and after the addition of the tholin sample (thin yellow bars), for the experiments in (a) pure $N_2$ plasma, and in (b) pure argon plasma. For (b), argon isotopes and double ionized ions are found at m/z 18, 19, 20, 36, 38, 40. Water traces are found at m/z 17, 18. Nitrogen traces at m/z 28. Plasma leads to the formation of traces of $H_2$ (m/z 2), CO (m/z 28) and $CO_2$ (m/z 44).

### 3.3- Variations with the plasma parameters in $N_2$-$H_2$

*3.3.1- Effect of the injected quantity of $H_2$*

To study more precisely the effect of the addition of $H_2$ in the gas phase, experiments with different amounts of $H_2$ (1%, 3% and 5%) were performed. To observe relative evolutions in the comparison of spectra, they are normalized over the cumulated intensities of all mass peaks. Figure 5 shows the spectra obtained when tholins are immersed into the plasma for different amounts of $H_2$ injected. The increasing amount of $H_2$ is easily observed at m/z 2.

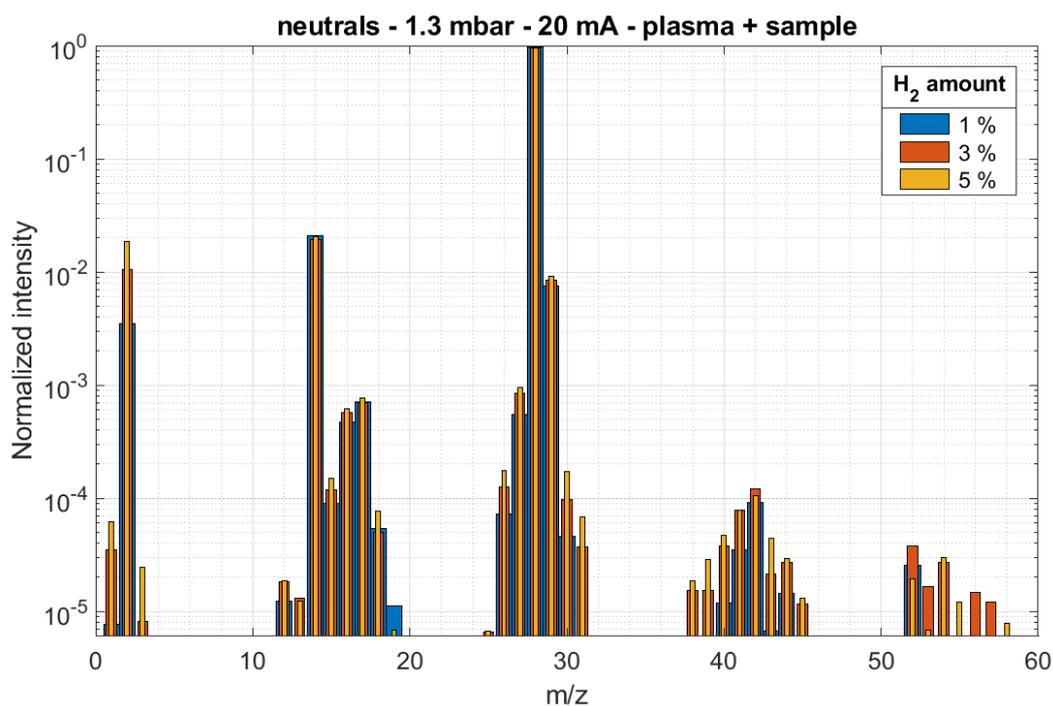

**Figure 5: Spectra obtained during the exposure of tholins to a $N_2$-$H_2$ plasma at different percentages of $H_2$. The MS is in the 'close' configuration.**

We observe that the quantity of ammonia at m/z 16 and 17 is rather constant with the injected $H_2$ percentage, from 1 to 5%. This trend is similar to the one observed in the plasma without tholins (Chatain et al., n.d.). On the other hand, Figure 6a-d-g shows that many carbon-containing species at m/z 27, 26, 30, 39, 40, 41 and 43 grow quasi-linearly with the amount of $H_2$% injected. The chemical pathways leading to their production depend directly on the hydrogen content in the gas phase.

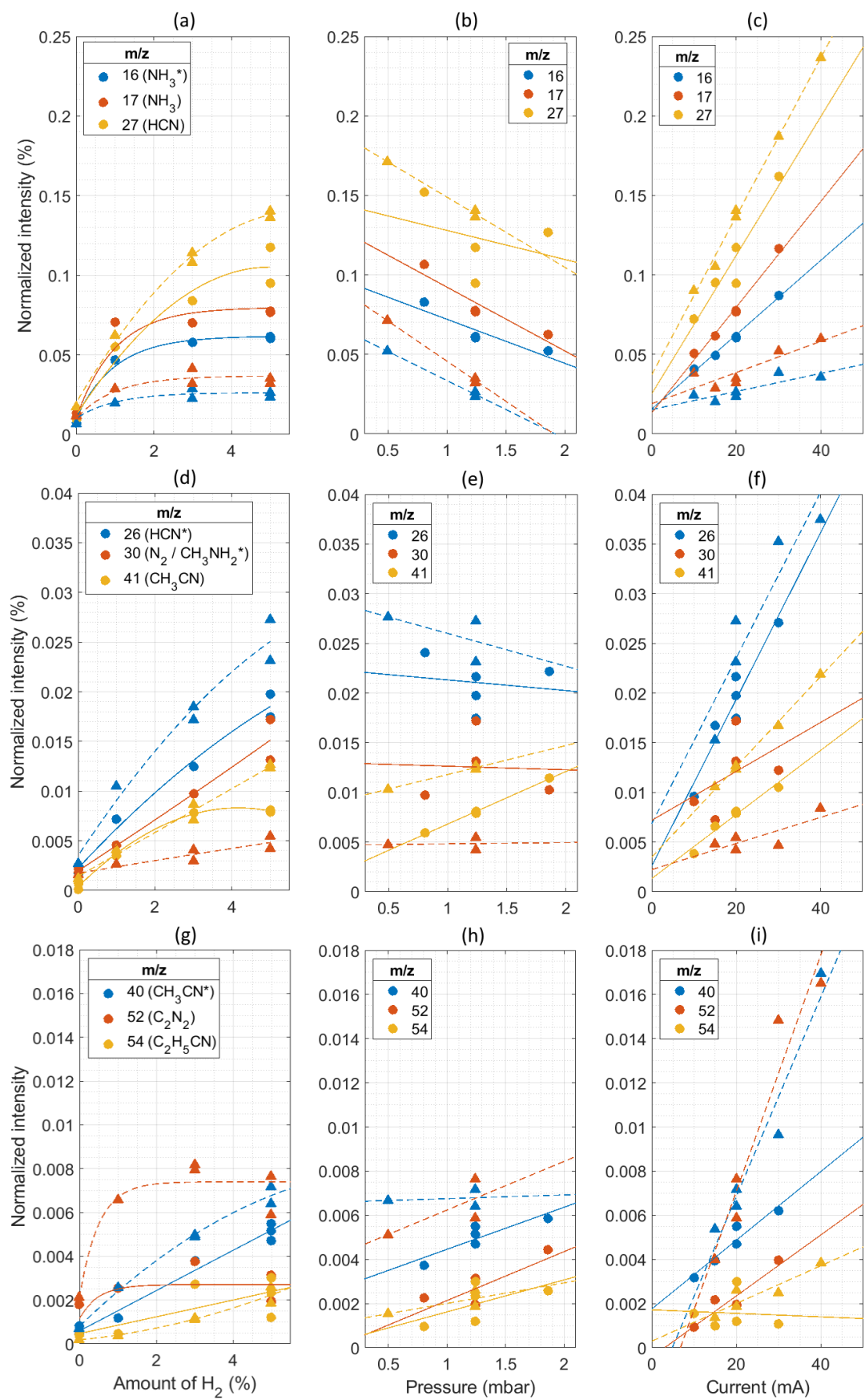

**Figure 6: Evolution of some masses with the H₂ amount, the pressure and the electric current, with reference conditions at 5% H₂, 1.3 mbar and 20 mA, in both the 'close' (circles) and the 'far' (triangles) configurations. The intensity is normalized to the sum of the intensities of all the masses. Suggested attributions are given in the first column. Lines are 1st and 2nd degree polynomial fits to guide the eye. * indicates fragments.**

*3.3.2- Observations with the variation of the pressure and the current*

A varying pressure has an effect on the production of the neutral species (see Figure 6b-e-h). The increase in pressure leads to a decrease of the ammonia proportion in the gas phase (at 16 and 17 u), while the ammonia partial pressure stays constant. The observations for HCN (at 27 u) are similar. On the other hand, the concentration of the carbon-containing species at 40 u, 41 u, 52 u and 54 u increase with pressure. A higher pressure favors the formation of heavier species. An increase of the electric current leads to a relative increase of many species, the main ones having peaks at m/z 16, 17, 26, 27, 38 to 41 and 52 (see Figure 6c-f-i). A higher current enhances the chemical reactions and the formation of the new neutral species. $C_2N_2$ at m/z 52 increases strongly especially in the configuration 'far'. These observations are discussed in detail in Section 3.4 and Section 5.1.3.

## 3.4- Conclusions on the neutrals formed by exposure of tholins to a $N_2$-$H_2$ plasma

The formation of the neutral products strongly depends on the plasma parameters. The same molecules are formed, but in varying quantities. This gives insights on the formation processes at stake. Results obtained in the previous section are synthesized in Table 3.

Ammonia partial pressure is globally constant with the $H_2$ amount and the pressure. These two parameters are therefore not the limiting variables to the production of ammonia. Besides, ammonia is detected in higher quantity when the MS stick is close to the plasma. Ammonia is certainly formed also on the MS stick surface. The limiting parameters for the production of ammonia are the presence of surfaces and the current. This property of ammonia to form easily on metallic surfaces is discussed in Cui et al. (2009) as it could also have happened on INMS walls on Titan. Due to this tendency to form on instrument walls or to stick on surfaces, ammonia is very difficult to quantify. Nevertheless, we show here that ammonia decreases strongly with the addition of tholins in the plasma. Ammonia or ammonia precursors are certainly destroyed by tholins or molecules formed from tholins.

HCN is the main product with the addition of tholins. It is likely to be formed at the surface of tholins, as it is the case for (C:H) polymers (see discussions above, i.e. Section 3.1). Its formation is enhanced with increasing $H_2$ amount and electric current. These results are consistent, as higher $H_2$% and current values increase the formation of H*, in great part responsible of the formation of HCN at the surface of tholins. With increasing pressure HCN proportion decreases, while products containing the function -C≡N increase (i.e. $CH_3$-CN, NC-CN, $NH_2$-CN, $C_2H_5$-CN). In conclusion, at higher pressure the formation of R-CN compounds is enhanced at the expense of the production of HCN in the gas phase.

The majority of the other detected molecules contains carbon and nitrogen, $C_xN_yH_z$. Their formation increase with the amount of $H_2$ in the plasma, which confirms that their formation pathways require H*, $H_2$, $H^+$ or H-containing species. Their production is also increased with an increasing electric current. This is especially the case for $CH_3$-CN and $C_2N_2$.

| molecule (m/z) | Effect of increasing H$_2$% | Effect of increasing pressure | Effect of increasing current | Evolution from the 'close' to the 'far' positions of the MS head | Previous work on Titan |
|---|---|---|---|---|---|
| NH$_3$ (16, **17**) | constant | relative decrease, but partial pressure constant | linear increase | loss | observations: (Cui et al., 2009; Magee et al., 2009; Yelle et al., 2010) models: (Vuitton et al., 2019) |
| HCN (26, **27**) | increase | relative decrease, but partial pressure quasi constant | linear increase | increase | observation: (Waite et al., 2005) models: (Vuitton et al., 2019) |
| CH$_3$-CN (38-**41**) | linear increase | linear increase | linear increase | increase | observation: (Cui et al., 2009) tholins: (He and Smith, 2014) models: (Balucani et al., 2012; Vuitton et al., 2019) |
| NC-CN (**52**) | increase | increase | linear increase | increase | observations: (Cui et al., 2009; Waite et al., 2005) models: (Vuitton et al., 2019) |
| CH$_3$-NH$_2$ (**30**-31) | linear increase | ~constant | ~constant | small decrease | tholins: (He and Smith, 2014) models: (Vuitton et al., 2019) |
| CH$_3$-CH=NH CH$_2$=N-CH$_3$ NH$_2$-CH=CH$_2$ *cyclic*-CH$_2$(NH)CH$_2$ (**43**) | increase | ~decrease | small increase or constant | small decrease | possible fragments of the structures proposed in (Pernot et al., 2010) model: (Balucani, 2012; Balucani et al., 2010) |
| C$_2$H$_5$-CN (**54**) | increase | ~increase | small increase or constant | - | observation: (Magee et al., 2009) models: (Vuitton et al., 2019) |

**Table 3: Global trends of the main products with the different plasma parameters. The last column refers to previous works done on the suggested molecules on Titan (observations, models or study of laboratory tholins).**

## 4- Evolution of positive ions

### 4.1- General observations

During the experiment, positive ions were also measured by mass spectrometry. Ions were hardly collected in the 'far' configuration at ~30 cm from the plasma. Therefore, all measurements presented have been done in the 'close' configuration, at a few centimeters from the glow discharge. Collected intensities are corrected as discussed in the Appendices D and E of Chatain et al. (2020b) to obtain the normalized ion flux. Ions are collected at a few centimeters from the plasma glow, and are certainly slightly different from the ions at the center of the glow, close to the sample.

Figure 7 presents the positive ion spectrum in the $N_2$-$H_2$ plasma, and its evolution with the insertion of the tholin sample. Many variations are observed, and in particular the decrease of m/z 16, 17 and 18, attributed mainly to the ammonia ions $NH_2^+$, $NH_3^+$ and $NH_4^+$. This is consistent with the decrease of ammonia discussed above. In addition, we observe the increase of some peaks and the apparition of new ones. It is especially the case of m/z 12, 13, 24 to 28, 36 to 41 and 52 to 54. These are carbon-containing species, thus formed from the interaction of the plasma with the tholins. Some attributions to these peaks are discussed in Section 4.4.

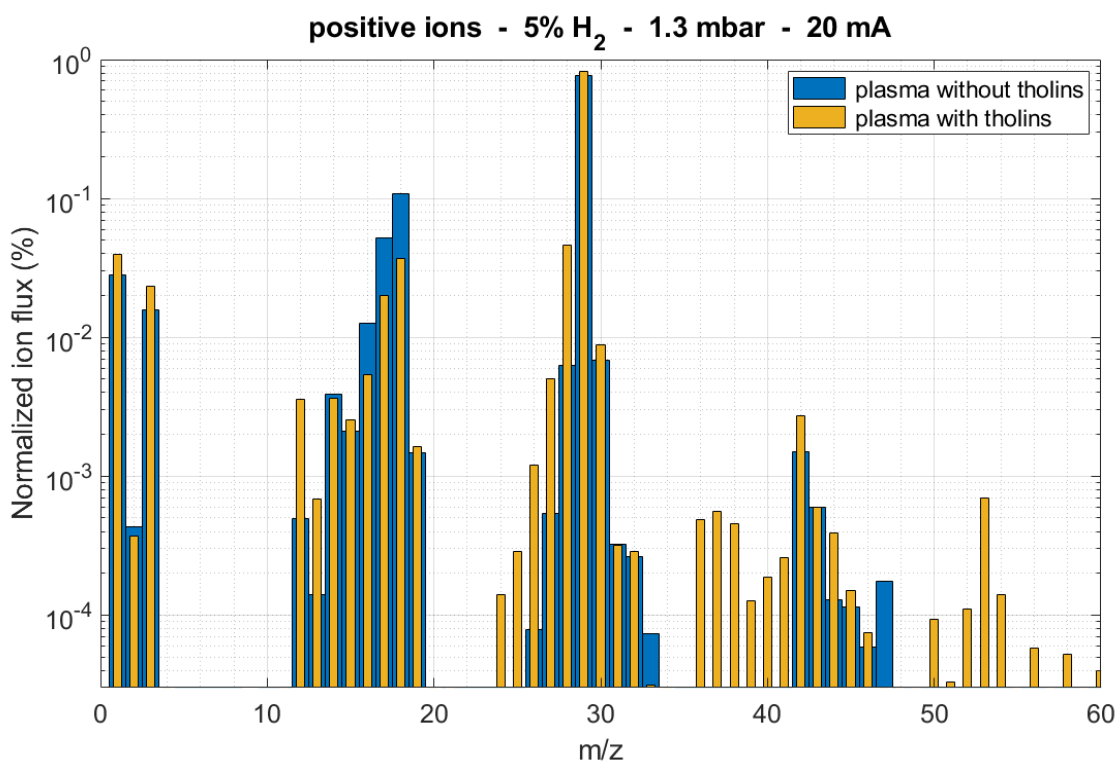

**Figure 7: Comparison of the positive ion spectra obtained in the reference experiment, in the $N_2$-$H_2$ plasma, and with the addition of tholins.**

An experiment done with a grid sample without tholins shows no variation to the $N_2$-$H_2$ plasma ion spectrum. The decrease of the ammonia ions and the apparition of the carbon-containing ions is entirely due to the presence of tholins. In addition, the reference experiment at 5% of injected $H_2$, 1.3 mbar and 20 mA has been done three times, with a good repeatability of the observed ion peaks. Figure A2-1 in Appendix 2 superimposes the three spectra, to give an idea of the variability of the spectrum for given plasma parameters.

### 4.2- Effect of the injected gas

Nearly no effect is observed in a pure $N_2$ plasma (see Figure 8a). The ignition of the plasma increases only slightly the amount of $H_3^+$, $C^+$, $C_2^+$, $CN^+$, $HCN^+$, $C_2N_2^+$. No variations are observed in argon plasma (see Figure 8b). These observations are consistent with the ones about neutrals. Therefore, the evolutions shown in Figure 7 require the presence of hydrogen and nitrogen in the gas phase. They are not due to direct decomposition by ion sputtering, similarly to the observations on the neutrals.

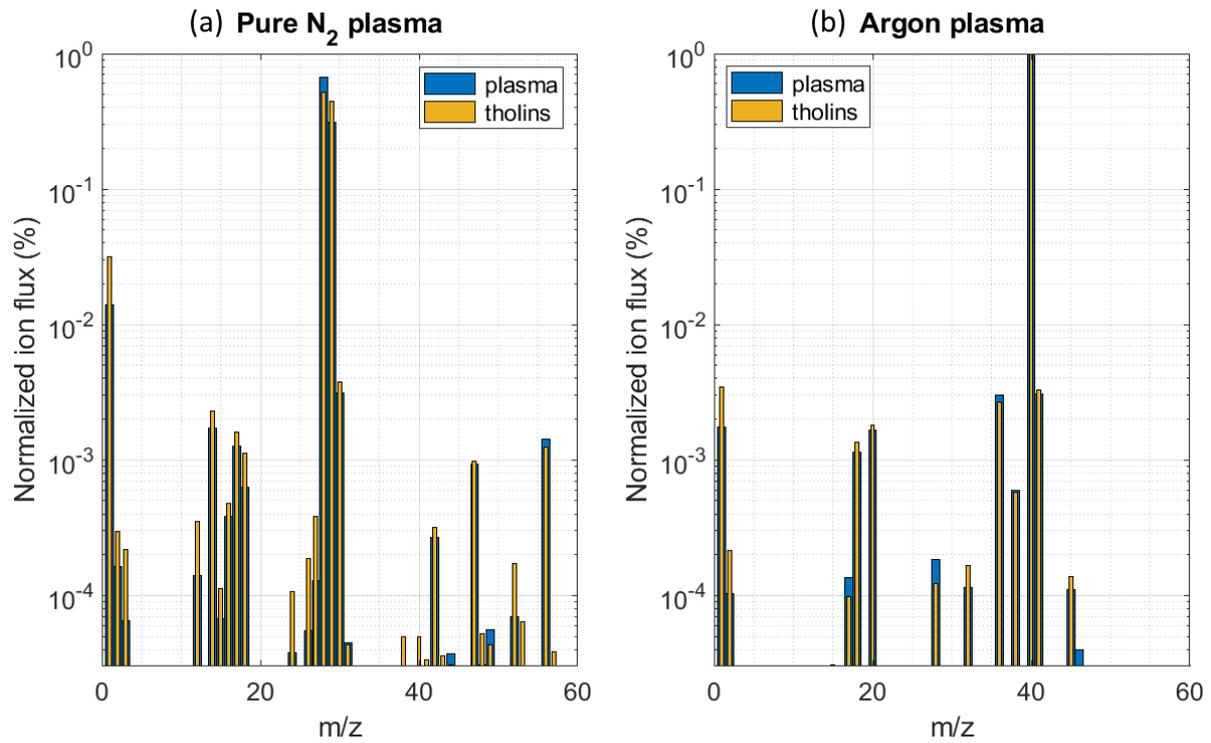

**Figure 8: Positive ion spectra obtained in (a) pure N$_2$ and (b) pure argon plasmas, and with the injection of tholins.**

### 4.3- Effect of the plasma parameters in N$_2$-H$_2$

Figure 9 shows the evolution of the ion spectrum with the H$_2$ percentage, the pressure and the current during the maximum erosion phase (3) introduced on Figure 2. More details on the major ions are given in Figure 10.

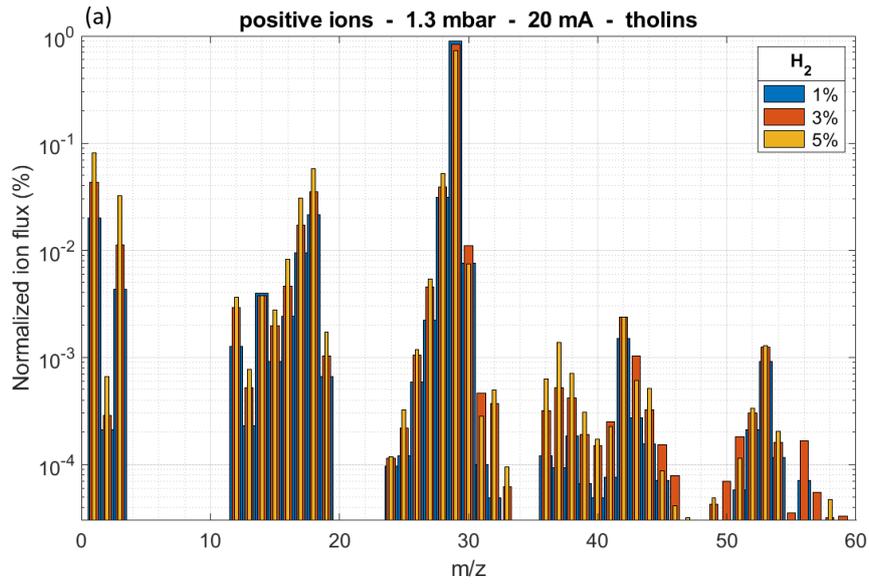
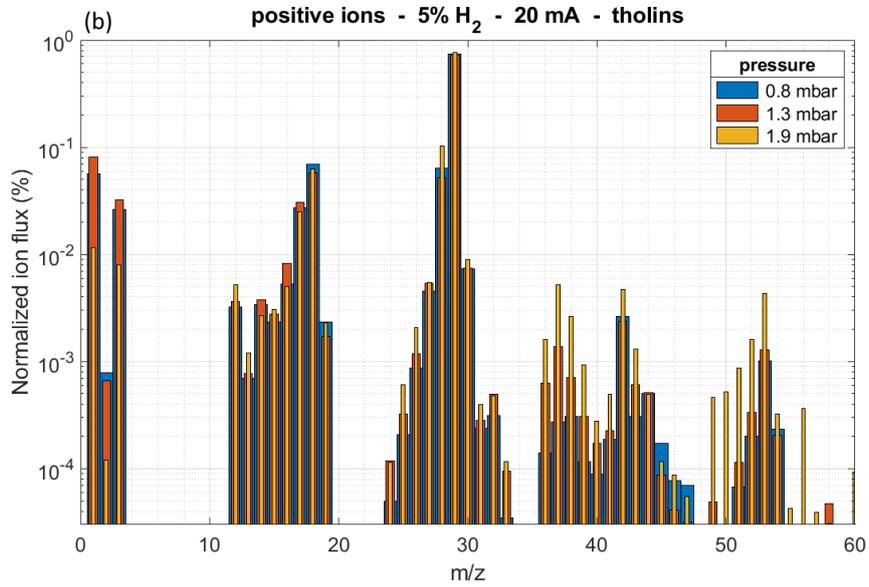
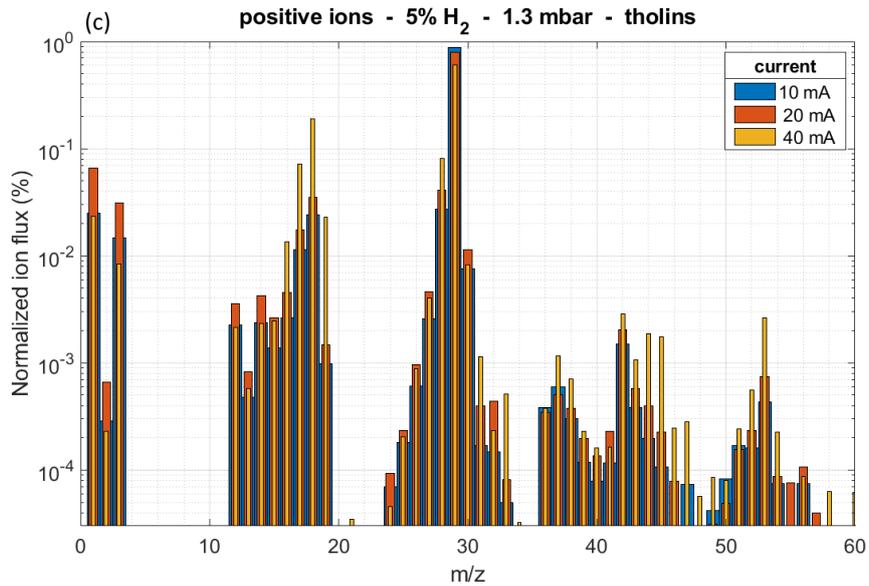

**Figure 9: Evolution of the positive ion spectrum in the presence of tholins with the increase of (a) the injected H₂ amount in the gas phase, (b) the pressure of the gas phase, and (c) the increase of the electric current.**

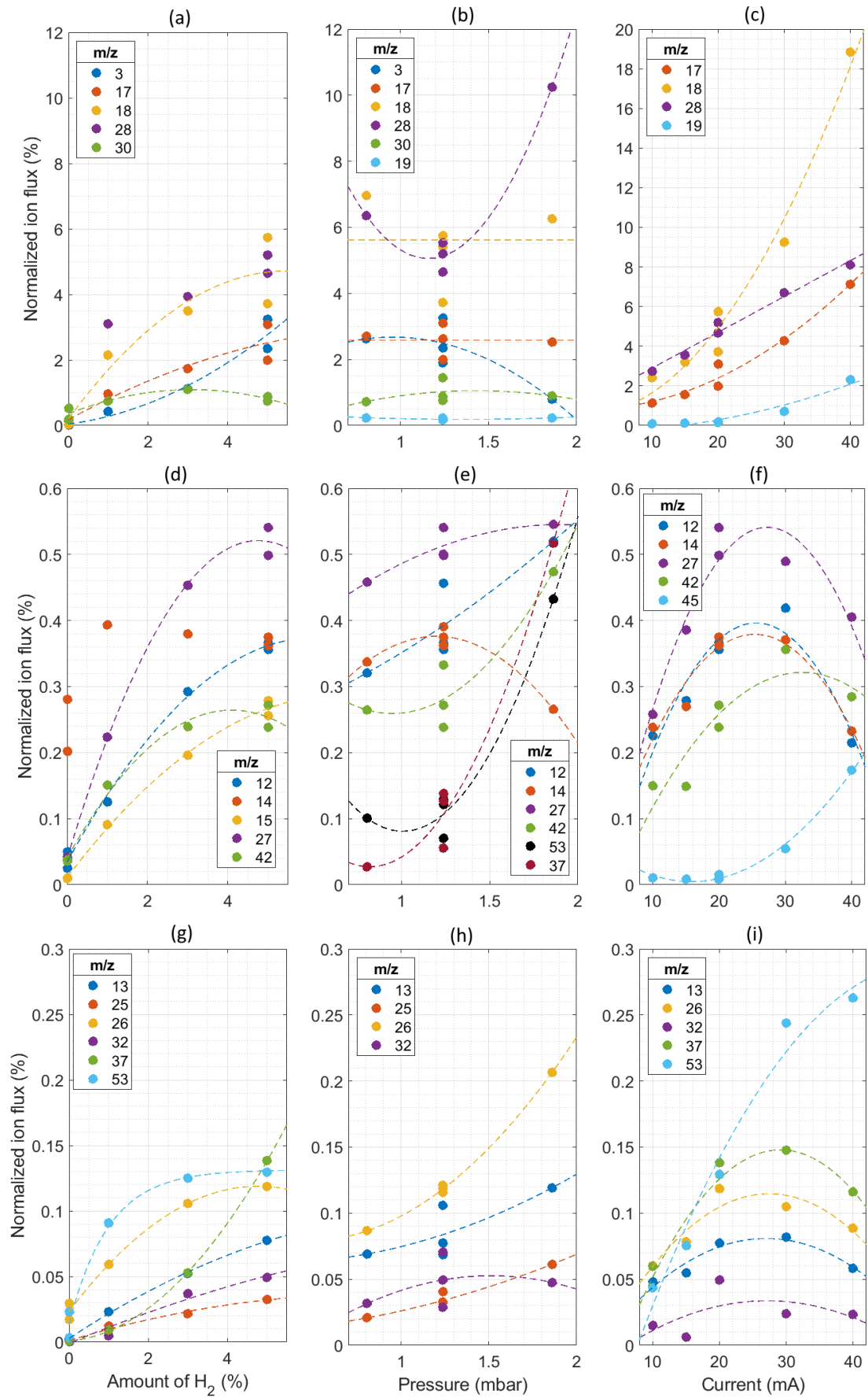

**Figure 10: Evolution of the main positive ions in the presence of tholins with the plasma parameters. The reference condition is at 5% H$_2$, 1.3 mbar and 20 mA. Ion attributions are suggested in the table of Section 4.4. Lines are 1st and 2nd degree polynomial fits to guide the eye.**

*4.3.1- Observations with the variation of the injected quantity of H$_2$ and the pressure*

Figure 9a shows the evolution of the ion spectrum in the presence of tholins with the increase of the injected H$_2$ amount in the gas phase. We observe that most of the ion species formed in the presence of tholins increase strongly. This confirms the fundamental role of hydrogen in the erosion of tholins. The heavier ions (m/z 42-43, 52-54) increase less than the lighter ones, especially m/z 3, 12-13, 15-18, 25-27 and 36-39. Ammonia ions (m/z 15-17) also increase, similarly to the case without tholins. Figure 9b shows the evolution of the ion spectrum in the presence of tholins with the increase of the pressure. This time, the increase mainly concerns the heavier ions: at m/z 36-39, 43, 49-53. Only a slight increase is observed at m/z 12-13 and 25-26. And similarly to the case without tholins (Chatain et al., n.d.), H$_3^+$ decreases and the ammonia ions stay constant. These observations are discussed in detail in Section 4.4 and Section 5.1.3.

*4.3.2- Effect of the current: fluorine contamination?*

Figure 9c shows the evolution of the ion spectrum in the presence of tholins with the increase of the electric current. Similarly to the case without tholins (Chatain et al., n.d.), it favors mainly the formation of the ammonia ions (m/z 16-18) and the formation of ions at m/z 19.

This peak at m/z 19 could be attributed to the enhanced desorption of protonated water. However, we suspect it comes from F$^+$ that could be formed at the higher current conditions by the sputtering of N$_2^+$ or N$_2$H$^+$ ions on the cathode (containing BaF$_2$) or the PTFE (polytetrafluoroethylene) junction present close to the MS collecting head. This observation can be linked to the detection of peaks at m/z 47, 66 and 85 in pure N$_2$ plasmas, that can be attributed to N$_2$F$^+$, N$_2$F$_2^+$ and N$_2$F$_3^+$. These three peaks are especially observed in the pure N$_2$ plasma case because the absence of hydrogen leads to a stronger N$_2^+$ sputtering on the cathode to obtain the same current. Therefore, we suspect that also in N$_2$-H$_2$ plasmas, fluorine-containing species could appear at the highest current conditions. In particular, N$_2$F$^+$/C$_2$H$_x$F$^+$ ions could explain the increase of m/z from 43 to 47 with the electric current, as well as NF$^+$/CH$_x$F$^+$ at 31-33 u and NH$_x$F$_2^+$/CH$_x$F$_2^+$ at 52 to 54 u. For precaution, experiments at high electric current are not investigated further. Remaining at low current also prevents gas heating (Chatain et al., n.d.).

## 4.4- Summary: positive ions formed by the exposure of tholins to a N$_2$-H$_2$ plasma

Table 4 collects all the observations presented above and associates possible ion attributions to each mass. These are discussed in the following paragraphs.

| m/z | plasma N$_2$ | plasma N$_2$-H$_2$ | N$_2$-H$_2$ + tholins | ↗ H$_2$ % | ↗ pressure | attribution |
|---|---|---|---|---|---|---|
| 2 | - | + | + | ↗↗ | ↘↘ | H$_2^+$ |
| 3 | - | ++ | ++ | ↗↗↗ | ↘↘ | H$_3^+$ |
| 12 | - | + | **++** | ↗↗ | ↗ | C$^+$ |
| 13 | - | - | **+** | ↗↗ | ↗ | CH$^+$ |
| 14 | + | ++ | ++ | - | - | **N$^+$** - CH$_2^+$ |
| 15 | - | + | + | ↗↗ | ↗ | NH$^+$ - **CH$_3^+$** |
| 16 | + | ++ | ++ | ↗↗ | - | **NH$_2^+$** - CH$_4^+$ |
| 17 | + | +++ | ++ | ↗↗ | - | **NH$_3^+$** - CH$_5^+$ |
| 18 | + | +++ | +++ | ↗↗ | - | **NH$_4^+$** - H$_2$O$^+$ |
| 19 | - | + | + | ↗↗ | - | H$_3$O$^+$ |
| 24 | - | - | - | - | ↗ | C$_2^+$ |
| 25 | - | - | - | ↗↗ | ↗↗ | C$_2$H$^+$ |

| m/z | plasma $N_2$ | plasma $N_2$-$H_2$ | $N_2$-$H_2$ + tholins | ↗ $H_2$ % | ↗ pressure | attribution |
|---|---|---|---|---|---|---|
| 26 | - | - | **+** | ↗↗ | ↗↗ | $CN^+$ - $C_2H_2^+$ |
| 27 | - | + | **++** | ↗↗ | ↗ | $HCN^+$ - $C_2H_3^+$ |
| 28 | +++ | ++ | **+++** | ↗ | ↗ | $N_2^+$ - **$HCNH^+$** - $C_2H_4^+$ |
| 29 | +++ | +++ | +++ | ↘ | - | **$N_2H^+$** - $CH_2NH^+$ - $C_2H_5^+$ |
| 30 | ++ | ++ | ++ | - | - | **$N_2H_2^+$** - isotope **$N_2H^+$**<br>$CH_2NH_2^+$ - $CH_3NH^+$ - $C_2H_6^+$ |
| 31 | - | - | - | ↗↗ | ↗ | $N_2H_3^+$ - $CH_3NH_2^+$ - $NOH^+$ |
| 32 | - | - | - | ↗↗ | ↗ | $N_2H_4^+$ - $CH_3NH_3^+$ - $NOH_2^+$ |
| 33 | - | - | - | ↗ | ↗↗ | $N_2H_5^+$ - $NOH_3^+$ |
| 36 | - | - | **+** | ↗↗↗ | ↗↗↗ | $C_3^+$ |
| 37 | - | - | **+** | ↗↗↗ | ↗↗↗ | $C_3H^+$ |
| 38 | - | - | **+** | ↗↗ | ↗↗↗ | $C_2N^+$ - $C_3H_2^+$ |
| 39 | - | - | - | ↗↗ | ↗↗↗ | $CHCN^+$ - $C_3H_3^+$ |
| 40 | - | - | - | ↗↗ | ↗↗ | $CHCNH^+$ - $CH_2CN^+$ - $C_3H_4^+$ |
| 41 | - | - | - | ↗ | ↗↗ | $CH_3CN^+$ - $C_3H_5^+$ |
| 42 | + | + | ++ | ↗ | - | $N_3^+$ - $N_2CH_2^+$ - $CH_3CNH^+$ - $C_3H_6^+$ |
| 43 | - | + | + | ↗ | ↗↗ | $N_3H^+$ - $N_2CH_3^+$ - $C_2H_3NH_2^+$ - $C_3H_7^+$ |
| 44 | - | - | **+** | ↗↗ | - | $N_3H_2^+$ - $C_2H_5NH^+$ - $C_3H_8^+$ |
| 45 | - | - | - | - | - | $N_3H_3^+$ - $C_2H_5NH_2^+$ - $C_3H_9^+$ - $CO_2H^+$ |
| 46 | - | - | - | - | - | $N_3H_4^+$ - $NO_2^+$ |
| 49 | - | - | - | - | ↗↗ | $C_4H^+$ |
| 50 | - | - | - | - | ↗↗ | $C_3N^+$ - $C_4H_2^+$ |
| 51 | - | - | - | ↗ | ↗↗↗ | $HC_3N^+$ - $C_4H_3^+$ |
| 52 | - | - | - | ↗ | ↗↗↗ | **$C_2N_2^+$** - $HC_3NH^+$ - $C_4H_4^+$ |
| 53 | - | - | **+** | ↗ | ↗↗↗ | **$C_2N_2H^+$** - $C_2H_3CN^+$ - $C_4H_5^+$ |
| 54 | - | - | - | ↗ | - | $CN_3^+$ - **$C_2N_2H_2^+$** - $C_2H_3CNH^+$ - $C_4H_6^+$ |

**Table 4: Positive ions peaks detected in $N_2$ and $N_2$-$H_2$ plasmas with tholins, their evolution with plasma parameters, and their suggested attributions. '+++' is for an ion flux > 3%, '++' for [0.3 ; 3]%, '+' for [0.03 ; 0.3]% and '-' for < 0.03%. '↗↗↗' is for x6+, '↗↗' for x3, '↗' for x2. Grey boxes show trends also observed in plasma without tholins. Red '+' in the fourth column indicates the ions strongly increased by the insertion of tholins (there are no tholins in the second and third columns). The main molecules suspected at a given mass are indicated in bold in the attribution column.**

### 4.4.1- $H^+$, $H_2^+$, $H_3^+$

The hydrogen ions are formed in the $N_2$-$H_2$ plasma. Variations with the plasma parameters ($H_2$ %, pressure and current) are similar in the plasma alone or with tholins. Consequently, the dominant processes dictating the hydrogen ions are those of a $N_2$-$H_2$ plasma discharge (see discussion in Chatain et al., n.d.).

### 4.4.2- 'C1 group': $NH_x^+$ and $CH_x^+$

$N^+$ and $NH_x^+$ ions are produced in the $N_2$-$H_2$ plasma. The addition of tholins in the plasma leads to a strong decrease of ions at m/z 16, 17 and 18 ($NH_2^+$, $NH_3^+$ and $NH_4^+$). Nevertheless, these masses are still dominant, and their variations with the injected $H_2$ amount, pressure and current is similar to the ones without tholins (Chatain et al., n.d.). This suggests that the peaks at m/z 16, 17 and 18 are still mainly due to $NH_2^+$, $NH_3^+$ and $NH_4^+$.

Nevertheless, new species also appear with the insertion of tholins. m/z 12 and 13 are attributed to $C^+$ and $CH^+$. They are strong indications of the presence of $CH_x^+$ ions. However, main ions as $CH_2^+$, $CH_4^+$ and $CH_5^+$ cannot be detected as the peaks at their respective masses are governed by $N^+$ (m/z 14), $NH_2^+$ (m/z 16) and $NH_3^+$ (m/z 17). On the other hand, as m/z 15 does not decrease with the insertion of tholins similarly to m/z 16, 17 and 18, we attribute it mainly to $CH_3^+$. Besides, its variation with plasma parameters is more similar to $C^+$ and $CH^+$ variations than those of m/z 16, 17 and $NH_4^+$.

The $NH_x^+$ ions variations are similar to those in plasma without tholins, but the variations of the $CH_x^+$ are different. While an increase in the hydrogen content increases the production of all the ions at m/z 12-13, 15-18, an increase of the pressure increases only the $CH_x^+$ ions. $CH_x^+$ ions are therefore more easily extracted from tholins at higher pressure compared to other ions. On the opposite, the increase of the current has mainly a strong effect on the ammonia ions. At higher currents, the production of the $NH_4^+$ ion is so efficient that it impacts strongly the total ion flux. This explains why the ratios of $CH_x^+$ decrease at higher current.

*4.4.3- 'C2 group': $N_2H_x^+$, $CNH_x^+$ and $C_2H_x^+$*

The insertion of tholins in the $N_2$-$H_2$ plasma leads to a strong increase in the ions from m/z 24 to 28. On the other hand, ions at m/z 30 to 33 do not change significantly. Besides, their evolution with plasma parameters are the same with and without tholins. This suggests that no main new species are formed from the interaction of the plasma with tholins at m/z 31 to 33. Nevertheless, peaks at m/z 29 and 30 are high, and could hide the formation of new ions.

The peaks at m/z 24 and 25 are attributed to $C_2^+$ and $C_2H^+$. This suggest the presence of others $C_2H_x^+$ ions, with x = 2 to 6. However, these ions have masses equal to other ions certainly present in large quantities: $CN^+$ at 26 u, $HCN^+$ at 27 u and $HCNH^+$ at 28 u. $N_2H^+$ prevents from detecting ions at m/z 29, as $C_2H_5^+$ or $CH_2NH^+$.

The peak at m/z 28 could also be attributed to $N_2^+$. Nevertheless, its variations with the plasma parameters are different from the case without tholins. It follows the variations of the peaks from m/z 24 to 27, sign that it is also created from the tholins. These carbon-containing species are produced with a higher efficiency if the injected $H_2$ amount and the pressure are increased. An increase followed by a decrease of the ion flux ratios are observed for the variations with the electric current. In conclusion, the variations with the hydrogen amount, the pressure and the current are the same as for the $CH_x^+$ ions.

*4.4.4- 'C3 group': $N_3H_x^+$, $N_2CH_3^+$, $C_2NH_x^+$ and $C_3H_x^+$*

New ion species appear between m/z 36 and 41 with the addition of tholins to the $N_2$-$H_2$ plasma. Peaks at m/z 36, 37 and 38 are particularly important and are the sign of the decomposition of the tholins in strongly unsaturated ions: $C_3^+$, $C_3H^+$ and $C_3H_2^+$. The last one can also be due to $C_2N^+$. The small peaks from 39 to 41 are certainly mainly due to the acetonitrile ions, $CHCN^+$, $CH_2CN^+$ and $CH_3CN^+$, as we observed previously the presence of acetonitrile. However, these masses can correspond also to $C_3H_x^+$ ions and to $CHCNH^+$. Similarly to the $CH_x^+$, $C_2H_x^+$ and $CNH_x^+$ ions, the ions at m/z 36 to 41 increase with the injected $H_2$ amount and the pressure.

Peaks at masses from 42 to 46 u are already present in the plasma without tholins (see Figure 7). They are certainly due to $N_3^+$ and $N_3H_x^+$ ions. Traces of $CO_2$, $O_2$ and carbon in the experiment can lead to small peaks at m/z 44 and 45 with $CO_2^+$ and $CO_2H^+$. In any case, with the insertion of tholins in the experiment, carbon-containing species may appear with masses between 42 and 46, in particular the ions linked to the neutrals observed at m/z 42 and 43 (see Section 3). Acetonitrile can also lead to the formation of $CH_3CNH^+$ at m/z 42. However, as the variations of these peaks with the plasma parameters is identical to the variations without tholins, we cannot conclude on the possible new carbon-containing species for these ions signatures between m/z 42 and 46.

*4.4.5- 'C4 group': $N_4^+$, $C_2N_2H_x^+$, $C_3NH_x^+$ and $C_4H_x^+$*

The addition of tholins into a $N_2$-$H_2$ plasma leads to a few heavier ions. The main one is at mass 53, surrounded by two smaller peaks at m/z 52 and 54. As $C_2N_2$ is a neutral detected in this experiment, we can legitimately expect the ions $C_2N_2H^+$ at 53 u, $C_2N_2^+$ at 52 u and $C_2N_2H_2^+$ at 54 u. Besides, these peaks increase with $H_2$ amount, pressure and power, similarly to the molecule $C_2N_2$.

An increase in pressure reveals ions at m/z 49 and 50, attributed to $C_4H^+$ and $C_4H_2^+$/$C_3N^+$. In the neutral mode, the detection of $C_2H_5CN$ suggests also the presence of related ions ($C_2H_xCNH_x^+$) at masses 53 to 56 in small quantities.

### 4.5- Discussion on the detected ions

Tanarro et al. (2007) studied the ions of a $H_2$-$N_2$-$CH_4$ DC glow discharge. They also observed $NH_x^+$, $CH_x^+$, $C_2H_x^+$, $HCNH^+$ and $CH_3CNH^+$ ions. The formation of these ions is directly linked to the addition or the loss of a proton to/from neutral molecules: $NH_3$, $CH_4$, $C_2H_2$, $C_2H_4$, $C_2H_6$, $HCN$ and $CH_3CN$. In our experiment, $CH_4$ is not injected in the gas phase, the $C_xH_y^+$ species come from fragments of tholins. In particular, $C_xH_y^+$ with smaller values for y are present in high quantity. Tanarro et al. (2007) suggest that their non-detection of CN and $C_2N_2$ is due to their high content in hydrogen (95%). On the opposite, $HCNH^+$ does not react with $H_2$ and $N_2$ and accumulate in their experiment. This can also explain the strong peak detected at m/z 28 in our experiment.

As previously discussed in the introduction, the etching of organic (C:H) films and polymers by $N_2$ and $N_2$-$H_2$ plasma discharges leads to the formation of new gaseous species. The main molecules detected are CN, HCN and $C_2N_2$ (Hong et al., 2002; Hong and Turban, 1999; Ishikawa et al., 2006; Kurihara et al., 2006), coming from the formation of the C≡N bond at the surface of the sample (Nagai et al., 2003, 2002). Van Laer et al. (2013) modelled the erosion of the films and predicted the following ions: $N_2H^+$, $N_2^+$, $H_2^+$, $NH^+$, $N^+$, $H^+$, $NH_2^+$, $NH_3^+$, $H_3^+$, $NH_4^+$, $CH_3^+$, $CH_4^+$, $HCN^+$, $CH_2^+$, $CH^+$, $C^+$. All those species are consistent with the observation of the erosion of tholins in THETIS. Nevertheless, heavier species are also detected here, as $C_3H_x^+$, $CH_3CNH^+$, $C_2N_2H^+$, which could come from fragments directly extracted from tholins.

## 5- Suggestion of surface processes and conclusions relevant for Titan

### 5.1- The surface processes in the experiment

Here we suggest some surface processes that could appear on tholins. They are strongly inspired by the surface model described by Van Laer et al. (2013), studying (C:H) films exposed to a $N_2$-$H_2$ plasma. We extended them to the case of tholins that contains also nitrogen. Figure 11 synthesizes these processes in a schematic.

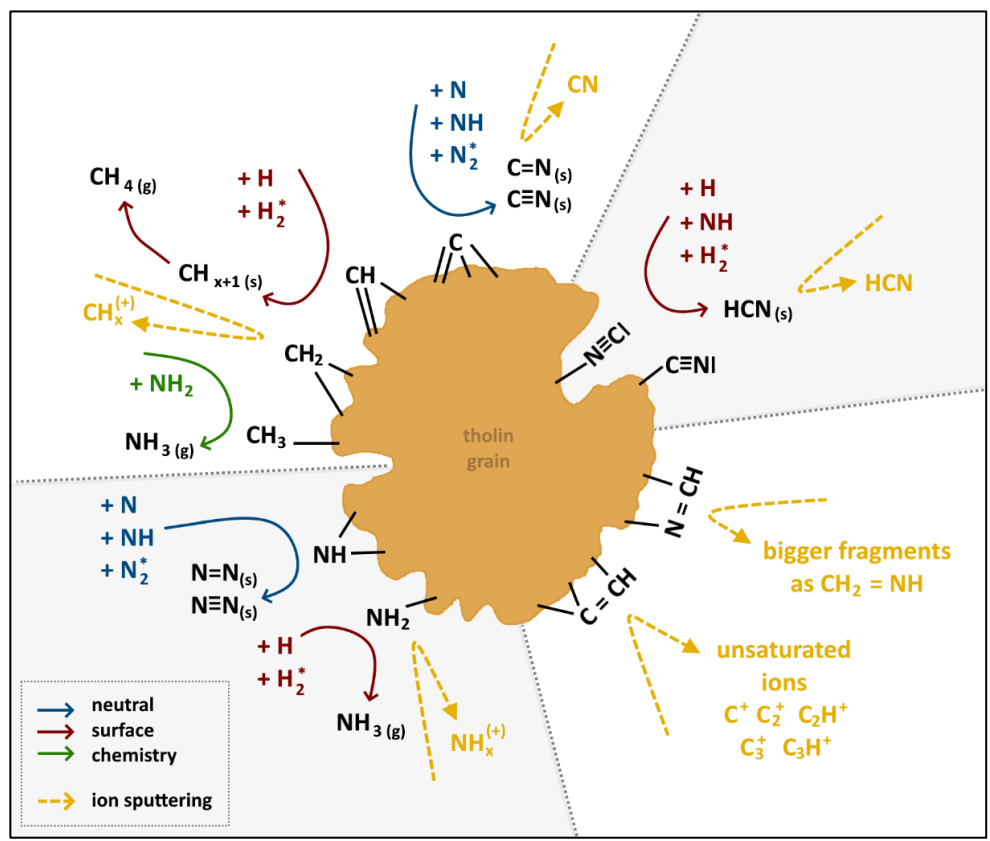

**Figure 11: Suggestion of surface processes on a tholin grain exposed to a $N_2$-$H_2$ plasma. The two upper parts come from Van Laer et al. (2013) and the two lower parts are suggestions inspired from THETIS observations.**

*5.1.1- Surface chemistry governed by radicals*

A first important process is the adsorption of neutral nitrogen on the surface. It comes mainly from the radical N, and NH and $N_2$ in a lower measure. Then, it reacts with the surrounding carbon and nitrogen atoms to form double and triple bonds. Studies of (C:H) films mention essentially C≡N bonds. However, in the case of tholins we can also consider N=N and N≡N bonds. Contrarily to NH, $NH_2$ is more likely to extract hydrogen from the tholin material to form the stable molecule $NH_3$. Neutrals containing hydrogen, namely H, NH and $H_2$, also attach to the surface, and hydrogenate the surrounding functions, in particular the unsaturated ones. Thus, it forms $CH_{x+1(s)}$ from $CH_{x(s)}$, up to the production of the stable molecule $CH_{4(g)}$. The hydrogenation of $C\equiv N_{(s)}$ gives $HCN_{(s)}$. Amines are also formed at the surface (Ishikawa et al., 2006).

Therefore, N, H and NH are actively used to form $HCN_{(s)}$, instead of $NH_3$ in the case without tholins. This could explain the decrease of ammonia in the gas phase at the insertion of the tholin sample. Nitrogen plasma are known for their massive production of metastable and vibrationally excited species (Guerra et al., 2004). Such species are reactive, though less than radicals, and present in large quantities. Consequently, they can also play a quantitative part in the surface chemistry described above.

*5.1.2- Ion sputtering to extract molecules from the tholins*

The surface chemistry described above leads in particular to the formation of $HCN_{(s)}$. However, if HCN is a stable molecule, it would not desorb without supplementary energy. The sputtering by heavy nitrogen ions ($N_2^+$ and $N_2H^+$) is required to desorb HCN. Otherwise, it would form a passivation layer at the surface of the sample grains. Sputtering by metastable states could also be possible.

Sputtering is necessary to erode the grains. However, the surface chemistry induced by the hydrogen does the up-front work. Indeed, it forms small stable molecules from the complex organic structure of the tholins. These small molecules are then easier to desorb by ion sputtering. This explains why only few molecules and ions are observed in the gas phase with the exposure of tholins to a pure $N_2$ plasma.

Unlike neutral species, ions are sensitive to the electric field in the sheath surrounding the tholins. They are strongly accelerated perpendicularly towards the surface. They bombard the surface and physically eject small pieces not strongly attached to the main structure of tholins.

Studies on (C:H) films observe the desorption of CN, HCN and $C_2N_2$. In the case of tholins, we observe many more molecules and ions. They could be formed afterwards in the gas phase. Nevertheless, most of the species in the gas phase are $N_2$-$H_2$ products and could not lead to the formation of complex organic species. Therefore, we suggest that at least part of the complex species observed are directly extracted from tholins. This is in particular the case of $CH_3CN$, $C_2N_2$, $CH_3NH_2$, the molecules suggested at m/z 43 and 54, and their corresponding ions. Besides, these molecules can be easily extracted from identified structures of tholin fragments analyzed by high-resolution mass spectrometry. In particular, He and Smith (2014) and Pernot et al. (2010) suggest amino-acetonitrile ($NC-CH_2-NH_2$) that could either give acetonitrile ($CH_3CN$) or methylamine ($CH_3NH_2$). Acetamidine ($CH_3-C(NH_2)=NH$) can lead to ethanimine ($CH_3-CH=NH$). m/z 44 could maybe also be due to $NH_2-CH=NH$ extracted from guanidine ($NH_2-C(NH_2)=NH$). Various molecules can similarly be extracted from cycles as methylimidazole and vinylimidazole.

THETIS ion spectra show the presence of strongly unsaturated ions, namely $C^+$, $CH^+$, $C_2^+$, $C_2H^+$, $C_3^+$, $C_3H^+$. These are probably directly extracted from tholins by the ion sputtering as secondary ions. They can perhaps correspond to the remaining parts of a cycle or a strongly unsaturated structure that has partly reacted with radicals in the processes presented above. As they are not observed in pure $N_2$ plasma exposure, we suspect that the hydrogen chemistry plays a role in the formation of these ions.

*5.1.3- Effect of the plasma parameters*

At higher pressure, absolute densities of reactive species increase (although their lifetime is shorter), and therefore the processes are accelerated. Besides, the reduced electric field E/N is smaller (Chatain et al., n.d.), which decreases the average electron energy, and therefore reduces the breaking of heavy tholin fragments in the gas phase by electrons. In addition, at higher pressure the sheath is more collisional and thus ions reaching the surface could be less energetic. The sputtering due to ions could consequently be less destructive. All of this could then explain the increased proportion of carbon-containing products, and especially the strong increase of the heavier ones.

As discussed above, $H_2$ plays a fundamental role in the chemical erosion of tholins. An increase of the hydrogen content between 0 to 5% therefore increases the hydrogen-induced chemistry at the surface of the tholins. It explains the increase of gas species produced from tholins, and tends to form smaller species.

*5.1.4- Comparison to FTIR results*

The processes described above are in agreement with the evolution of the IR absorption spectrum of tholins studied with the same setup in Chatain et al. (2020a).

First, as observed in Chatain et al. (2020a) by the SEM images and the absolute absorbance, the tholins on the sample are physically eroded and removed from the pellet. This loss is realized by the combined effects of the chemical surface alteration by the radicals, and the ion sputtering.

Secondly, the chemical evolutions observed by IR absorption of the solid phase are consistent with the chemical processes happening at the surface of grains. The bands of nitriles around 2100-2250 $cm^{-1}$ (4.76-4.44 µm) are easily extracted in the gas phase as HCN, explaining their faster erosion than the other bands. Besides, isonitriles (2135 $cm^{-1}$/4.68 µm), the most unstable nitriles are lost in the first place. The study of the IR absorption also showed the apparition of a new nitrile band at 2210 $cm^{-1}$ (4.52 µm), which could be attributed to the new C≡N bonds forming at the surface of the tholins, with the processes presented just above. The decrease of C=N bonds (1665 $cm^{-1}$/6.01 µm) and of $CH_3$ bonds possibly linked to >CH-CN (2950 $cm^{-1}$/3.39 µm) are also consistent with the processes discussed above, in which C=N are transformed into C≡N bands and HCN and $CH_3$-CN desorbed into the gas phase.

## 5.2- Conclusions relevant for Titan

*5.2.1- Radicals in Titan's ionosphere*

Radicals and excited species are also present in quantity in planetary ionospheres (Majeed et al., 2004, 1991). Nevertheless, they are complicated to measure and there had been only few detections. However, models of the upper atmosphere, especially of Titan, suggest that radicals and excited species play an important role in the ionospheric chemistry (Vuitton et al., 2019). Lavvas et al. (2011a) studied the formation of the main radicals and excited species in Titan ionosphere, including photoionization of $H_2$ and $N_2$. In particular, between 1000 and 1200 km, the radical H is present at 0.1%, N at 0.01%, and NH and $NH_2$ at a ratio of $10^{-6}$.

Therefore, atomic H is present in quantity in the ionosphere. Sekine et al. (2008a, 2008b) studied the interaction between atomic H and tholins. They observed a strong loss of H atoms by heterogeneous processes, mainly from hydrogenation and $H_2$ recombination, and only slightly from chemical erosion. They suspect that 60 to 75% of hydrogen atoms are removed from the gas phase by these processes in the stratosphere and the mesosphere. Contrarily to the hydrogenation process that adds an H atom in the structure of the tholins, the $H_2$ recombination removes a H atom present in the tholins structure. Therefore, it enables also the formation of unsaturated structures. We went further with the THETIS experiment, by adding N-bearing radicals and ions to the gas phase, and found that they work in close combination with the H radicals at the surface of tholins.

Lavvas et al. (2011b) also observed the importance of radical-including chemistry in the formation of the aerosols. However, they focused on 'heavy' radicals, as $C_2H$, CN and HCCN. Such radicals lead to the formation of heavier species in the gas phase that ultimately adsorb on aerosols, or directly adsorb themselves on the aerosols.

*5.2.2- Heterogeneous chemistry on the aerosols induced by H, N radicals, and collision with energetic species*

As H and N radicals, as well as others $N_2$, $H_2$, $N_xH_y$ radicals or excited species, are strongly represented in Titan's ionosphere, we suspect that the processes described in Section 5.1 could also happen up to a certain extent on Titan's ionospheric aerosols. In particular, we are able to make two suggestions.

First, the H and N radicals could induce a chemistry at the surface of the aerosols. N would lead to the addition of nitrogen at the surface that would evolve into double and triple bonds between N and an atom of the aerosol structure, in particular forming C≡N. It would produce new unsaturated structures at the surface of the aerosol. On the other side, H adsorption would hydrogenate some other parts of the surface. Then, aerosol structure would continuously evolve during its growth in the ionosphere.

Secondly, the aerosols could also exchange small fragments with the surrounding gas phase. Sputtering induced by collision with energetic species could desorb small molecules or ions from their surface. It could especially lead to HCN and R-CN gas species.

Quantitatively, the pressure in Titan's ionosphere is lower than in the THETIS experiment. However, the electron density measurements in Chatain et al. (n.d.) shows that the ionization degree is similar. This suggests a similar equilibrium between the chemistry induced by neutrals and ions. The sheath induced by the grains on Titan is not collisional at such low pressures. The acceleration of ions toward the surface depends mainly on the charge of the aerosols, which is likely to evolve with altitude (Shebanits et al., 2016). It would be interesting to model the energy of the incoming ions on Titan's aerosols and compare it to laboratory values. Concerning radicals, estimations in THETIS plasma conditions give radical mixing ratios about one order of magnitude higher than the model results on Titan (~1% for H and N, ~$10^{-4}$ for NH and ~$10^{-5}$ for $NH_2$) (V. Guerra, *private communication*). These values will have to be experimentally measured in THETIS. Nevertheless, higher radical ratios (x10) would increase the surface chemistry induced by radicals on the grains in THETIS compared to Titan.

The THETIS experiment shows the increase in the production of tholin-extracted species in the gas phase with increasing pressure and hydrogen content. Both variations are anti-correlated on Titan, with the increase of hydrogen at higher altitudes, while pressure decreases. However, in THETIS we observed heavier molecules desorbed with increasing pressure and decreasing $H_2$ amount. This suggests a desorption of heavier molecules at lower altitude on Titan.

*5.2.3- Comparison to INMS spectrum*

The ions observed in THETIS have all been detected in Titan's ionosphere by the mass spectrometer of Cassini, INMS (Magee et al., 2009; Mandt et al., 2012; Waite et al., 2007). This confirms that the processes going on in THETIS could also be part of the complex Titan chemistry. Figure 12 shows a comparison of a THETIS spectrum to a spectrum by INMS at 1016 km on Titan during the T40 flyby.

The chemistry occurring in Titan's ionosphere is very complex. Models up to now essentially focused on the gas phase chemistry. However, a strong disagreement between model results and Cassini data on nitriles led the latest models to take into account the sticking of the polar molecules (nitriles) and of H atoms onto the aerosols. Aerosols are used as a sink of molecules/atoms. Nevertheless, no model investigated the evolution of the chemical structure of the aerosols after the adsorption of these species. They do not either include the possibility of aerosols to give back small organic molecules in the gas phase. A further investigation on such heterogeneous processes happening at the surface of aerosols could bring some insights, especially concerning the HCN and CN-bearing molecules present in Titan's ionosphere.

Besides, the main ion peak detected by INMS is attributed to $HCNH^+$. Models predict that $HCNH^+$ comes mainly from the protonation of HCN. $HCNH^+$ is an ion of fundamental importance in the complex chemistry going on in Titan's ionosphere. We observe also a strong production of HCN, and $HCNH^+$ in the THETIS experiment, with the exposure of tholins to $N_2$-$H_2$ plasma species. As both $N_2$-$H_2$ plasma species and organic aerosols are present in Titan's ionosphere, we suspect that the processes leading to the formation of CN and HCN at the surface of aerosols could also happen on Titan. HCNH$^+$ is an important species on Titan and its production is not yet well constrained (Vuitton et al., 2019; Westlake et al., 2012), therefore it is worth investigating these new heterogeneous formation processes.

Apart from HCNH$^+$, the THETIS experiment leads also to the formation of acetonitrile, cyanogen and their protonated ions: $CH_3CNH^+$ (m/z 42) and $C_2N_2H^+$ (m/z 53). These molecules and ions are also found in high quantities on Titan. This suggests that they could be bricks forming the aerosols and ejected back in the gas phase after ion sputtering.

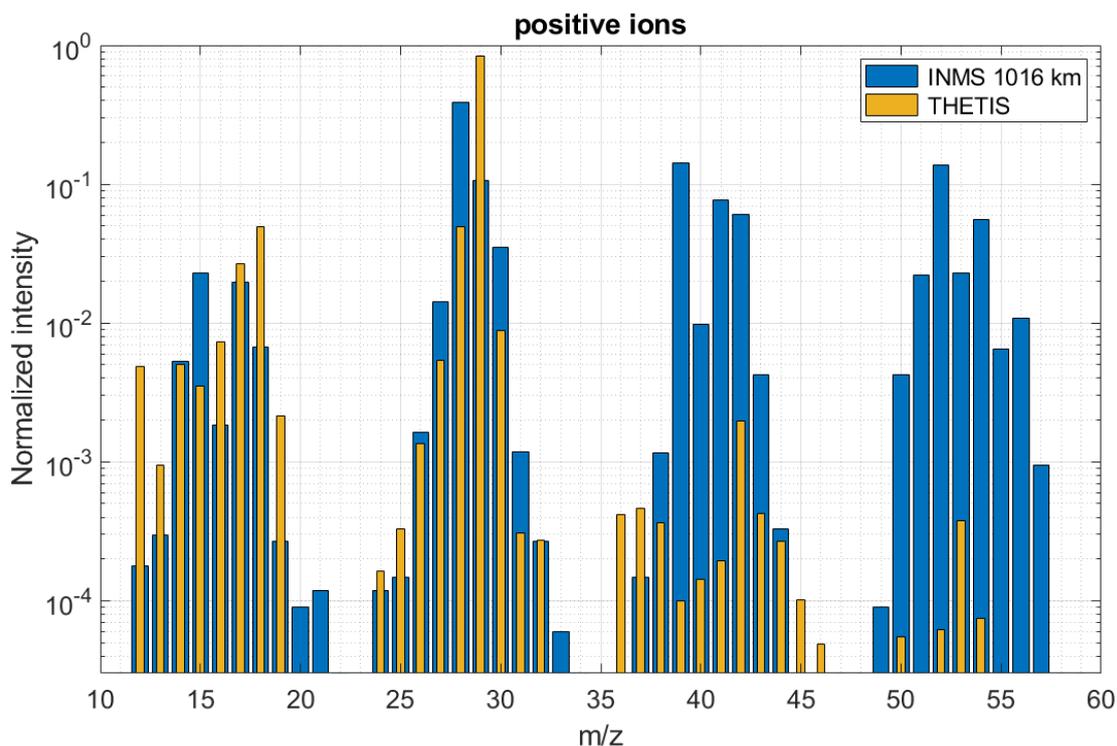

**Figure 12:** Comparison of the THETIS ion spectrum in reference conditions (5% $H_2$, 1.3 mbar, 20 mA, tholins) to a spectrum taken by INMS in Titan's ionosphere on the T40 flyby (Magee et al., 2009). Note: the THETIS spectrum only aims to reproduce the ions formed by interaction of $N_2$-$H_2$ plasma species with aerosols. INMS also includes the products of the complex $N_2$-$CH_4$ ion chemistry in the ionosphere. Consequently, intensities should not be directly compared: THETIS ions only give an idea of which ions in Titan's ionosphere could partly come from the plasma-aerosols interaction.

## 6- Conclusion

The THETIS experiment has been developed to study the interactions between analogues of Titan's aerosols (tholins) and the erosive $N_2$-$H_2$ plasma species found in Titan's ionosphere. Following a first paper on the investigation of the evolution of the solid phase by Scanning Electron Microscopy and IR transmission spectroscopy (Chatain et al., 2020a), this second paper focuses on the gas phase, studied by neutral and ion mass spectrometry of the gas phase.

We observe that newly formed HCN, $NH_3$-CN and $C_2N_2$ are extracted in quantity from the tholins as well as some other carbon-containing species and their derived ions. On the other hand, the production of ammonia is strongly decreased, certainly because the H, NH and N radicals are rather used for the production of HCN at the surface of tholins. In the last section we suggest heterogeneous processes based on our experimental observations, on the previous results on the solid phase in Chatain et al. (2020a) and on previous works on $N_2$-$H_2$ plasma etching on organic films. In brief, chemical processes induced by radicals at the surface modifies and weakens the tholin structure, and ion sputtering desorbs small molecules and highly unsaturated ions.

On Titan, the heterogeneous chemistry could similarly participate to the formation of C≡N bonds in the aerosol structure, as well as the production of HCN or R-CN species in the gas phase. These evolutions induced by the presence of $N_2$-$H_2$ plasma species in Titan's ionosphere are in competition with carbon-growth processes led by the presence of methane. The competition between $N_2$-$H_2$ erosion processes and carbon-growth processes will be investigated in future works.

## Appendix 1- More details on the neutral MS results

### A1.1- Repeatability of the results

The experiment presented on Figure 3 in section 3.1.3 has been repeated three times with exactly the same conditions and on different weeks. Results are given on Figure A1-1. Results are highly repeatable.

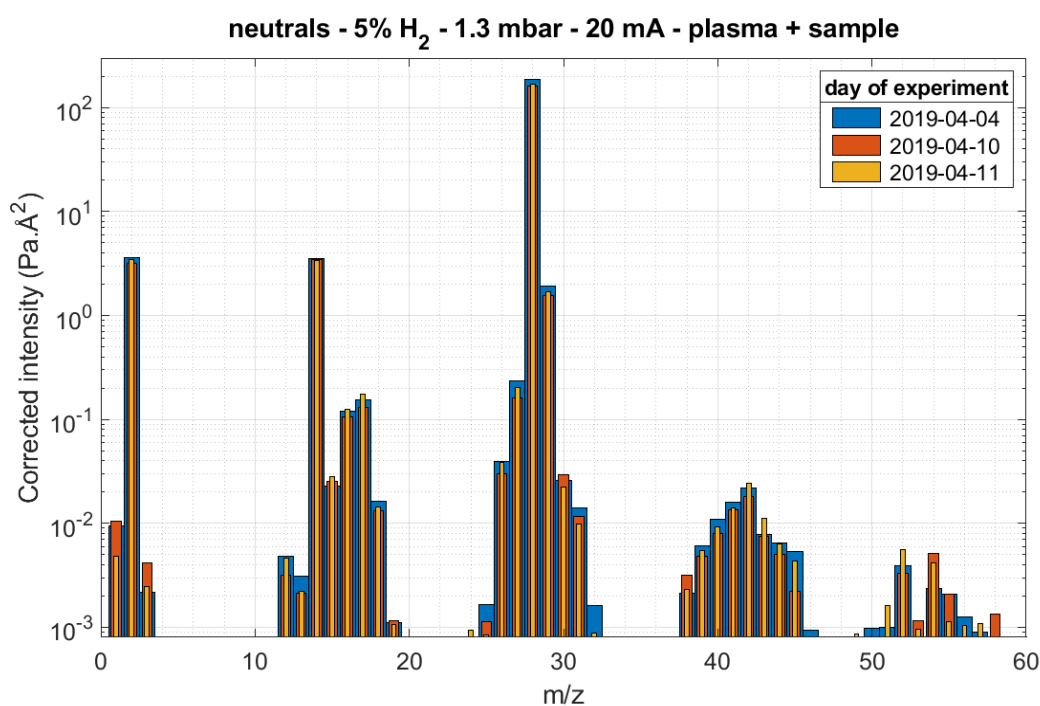

**Figure A1-1: Repeatability of the reference experiment on three different days. Situation with the sample added to the plasma.** Figure 3 corresponds to day 2019-04-10.

### A1.2- Effect of the position of the MS collecting head

The results presented in Figure 3 and Figure A1-1 were obtained with the MS collecting head aligned with the plasma axis, as shown in the 'close configuration' in Figure 1b and Figure A1-2. In this configuration, the plasma is only a few centimeters away from the MS. Then, the metallic collecting head is very close to the plasma and could induce some perturbations.

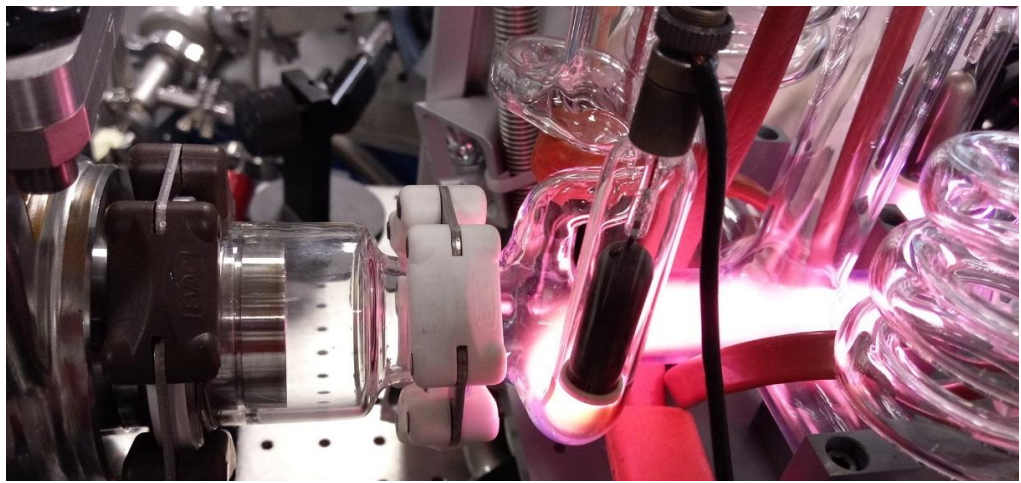

**Figure A1-2: Configuration with the MS collecting head close to the plasma (~5 cm).**

To quantify this effect, two other experiments were conducted in the same reference conditions, but with the MS collecting head further from the plasma (at ~30 cm). It is the 'far' configuration presented on Figure 1b. The results (shown in Figure A1-3) are globally identical. The main observation in the 'far' case is that ammonia is lost in greater quantity after the insertion of tholins. The higher level of ammonia detected in the case closer to the plasma suggests that it could be formed on the collecting head even if not directly in contact but close to the plasma. Indeed, ammonia is mainly formed by radical species that have a longer life time than excited species, and can easily reach the surface of the MS stick. We notice that in the case of a $N_2$-$H_2$ plasma without tholins (Chatain et al., n.d.), the detection of ammonia does not depend on the distance of the MS stick to the plasma. The insertion of tholins disturbs the production of ammonia and certainly changes its production processes.

Similarly to the $NH_3$ modifications, peaks at m/z 30 and 43 proportionally increase slightly in the 'close' configuration (see Figure A1-3). According to Table 2, both molecules certainly contain amines. On the opposite, -C≡N containing species are present in slightly higher proportions in the 'far' configuration: HCN (26-27 u), $CH_3$-CN (40-41 u) and $C_2N_2$ (52 u). In the 'close' configuration, molecules produced on the tholins have to go through 7-8 cm of plasma before reaching the MS. Contrarily, in the 'far' configuration they only have a few millimeters of plasma to cross. In this configuration, HCN, $CH_3$-CN and $C_2N_2$ are certainly less altered by the plasma in the gas phase.

Ion measurements (see Section 4) can only be done in the 'close' configuration. Therefore, each of the two configurations ('close' and 'far') has its advantages and drawbacks: consistency with ion measurements versus ammonia surface catalysis. To validate the results presented further, experiments have been performed at least once in each position. We observe globally the same trends in the two configurations with the variation of plasma parameters.

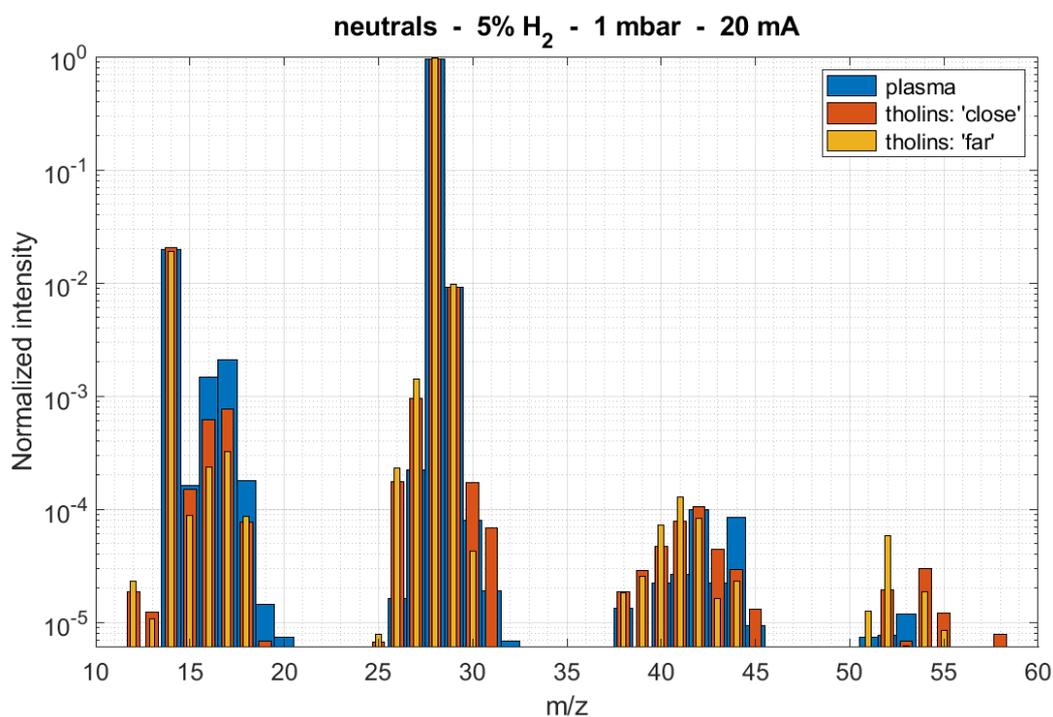

**Figure A1-3:** Comparison of the spectrum acquired without tholins, identical in the configurations 'close' and 'far' (large blue bars), with the spectra acquired after the insertion of tholins in the configurations 'close' (medium orange bars), and 'far' (thin yellow bars).

### A1.3- Test of a grid sample without tholins

An experiment was done without tholins on the grid. In this case, no changes have been detected with the addition of the empty grid sample, except perhaps a very slight increase of ammonia (x2), as shown in Figure A1-4. The metallic grid certainly enhances the surface production of $NH_3$.

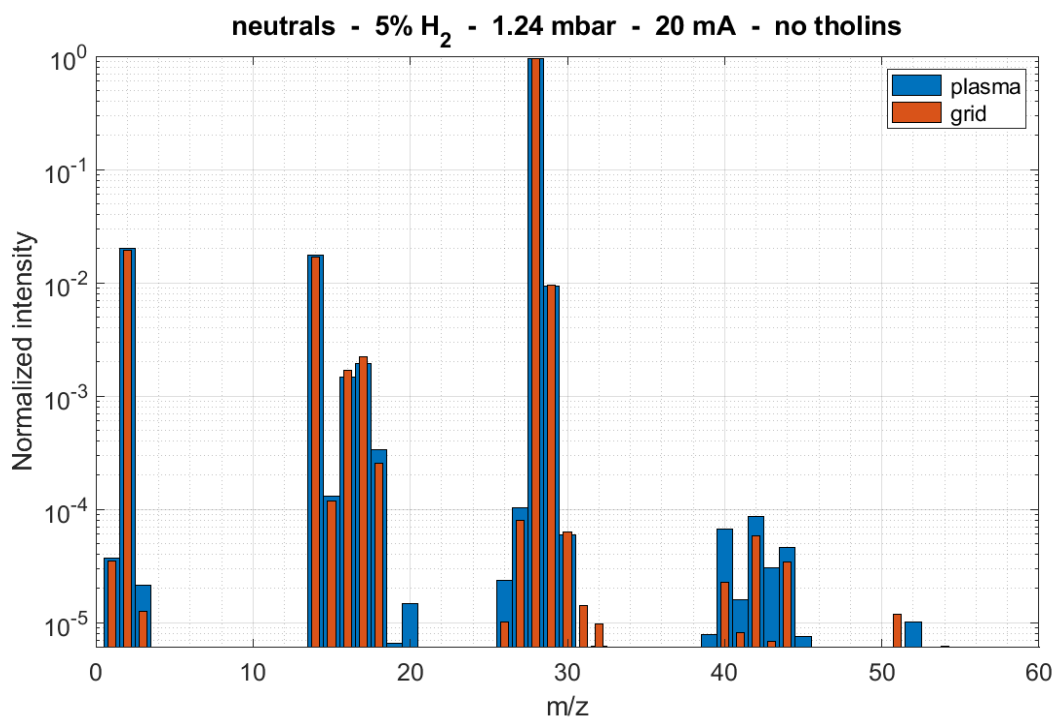

**Figure A1-4: Spectra obtained for the experiment in reference conditions in the 'close' configuration. Comparison of the spectrum acquired without tholins (large blue bars), with the spectrum acquired after the insertion of a grid without tholins (medium orange bars).**

## Appendix 2- More details on the ion MS results

The reference experiment at 5% of injected $H_2$, 1.3 mbar and 20 mA (detailed in Section 4.1) has been done three times. Figure A2-1 superimposes the three spectra, to give an idea of the variability of the spectrum for given plasma parameters.

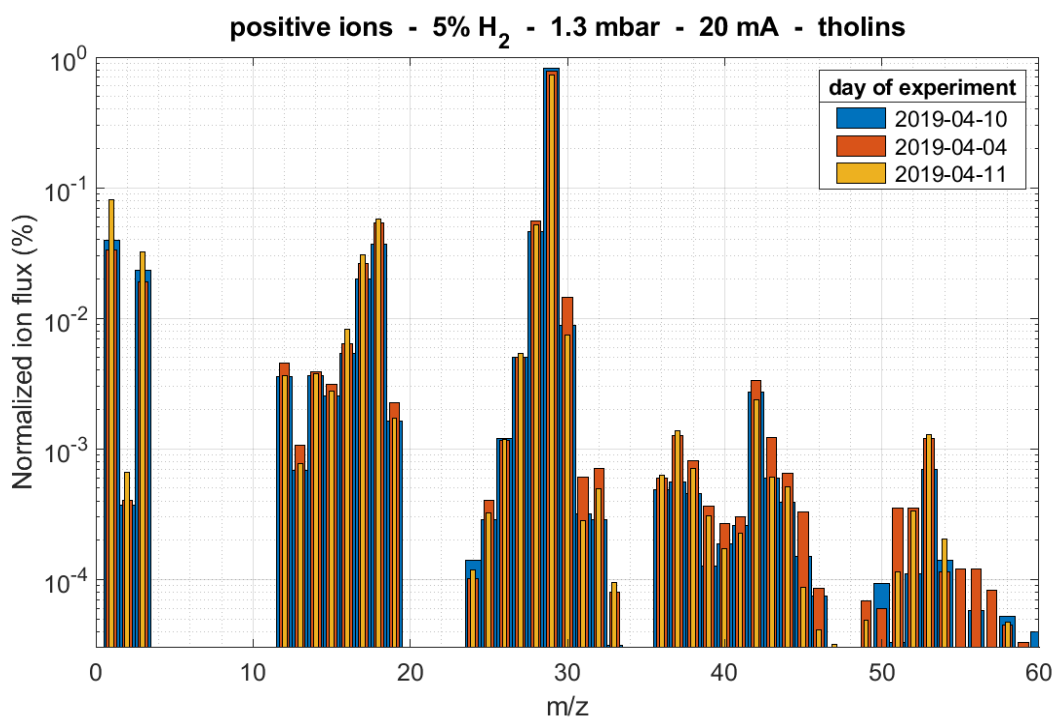

**Figure A2-1: Repeatability of the positive ion spectra obtained in the reference experiment with tholins.**

## Data availability

Raw data from mass spectrometry measurements, Matlab codes to process the data, processed Matlab variables, Matlab codes to plot the data and Matlab figures corresponding to the figures in the paper are available on Zenodo at the DOI: 10.5281/zenodo.7331908.

## Acknowledgments

NC acknowledges the financial support of the European Research Council (ERC Starting Grant PRIMCHEM, Grant agreement no. 636829). AC acknowledges ENS Paris-Saclay Doctoral Program. AC thanks Guy Cernogora for insightful discussions on the DC glow discharge.